\documentstyle [11pt]{article}
\begin{document}
\title{\bf The Theory Of Relativity - Galileo's Child}
\author{Mitchell J. Feigenbaum \\The Rockefeller University}
\date{May 25, 2008}
\maketitle

\begin{abstract}
\noindent We determine the Lorentz transformations and the kinematic content and dynamical framework of special relativity as purely
an extension of Galileo's thoughts. No reference to light is ever required: The theories of relativity are logically independent of
any properties of light. The thoughts of Galileo are fully realized in a system of Lorentz transformations with a parameter $1/c^2$,
some undetermined, universal constant of nature; and are realizable in no other. Isotropy of space plays a deep and pivotal role in
all of this, since here three-dimensional space appears at first blush, and persists until the conclusion:  Relativity can never
correctly be fully developed in just one spatial dimension.
\end{abstract}

%%%%%%%%%%%%%%%%%%%%%%%%%%%%%%%%%%%%%%%%%%%%%%%%%%%%%%%%

\section{Introduction}

The theory of inertial, uniform motions is a theory of, and the world of Galileo. That these motions have a relative meaning only,
``the principle of relativity", was again, the conception of Galileo. My purpose here is not hagiography - although there is also a
palpable ``villain". These thoughts, nearing four hundred years old, are potent and, when fully mathematically developed,
extraordinary fecund. It is my purpose here to show precisely that in these seminal concepts of Galileo lies the special theory of
relativity with no further additional physical insight or knowledge. In particular, it shall transpire that the physical phenomenon
and properties of light are logically independent of the present theories of relativity. (With special relativity in the tangent
spaces, general relativity is determined as well.) Apart from some historical commentary at the end, this paper is purely devoted to
an epistemologically correct determination of this theory, carefully excising all
that is superfluous and redundant.\\

The worlds of Galileo are identical environments, but moving uniformly, rectilinearly with respect to one another. In each, all
observed properties of the world are, to their respective denizens, identical. This is one of two arguments Galileo adduced for
inertia. It is most important to realize - as Galileo intended his intelligent reader to conclude - that this \textit{is} a full
statement about inertia. The pictorially rich exposition in ``Dialogue Concerning the Two Chief World Systems" determines that an
arbitrarily smooth and round ball sitting on an arbitrarily smooth table in the large aft chamber of a great ship, silently moving
under a uniform wind, simply reposes itself at rest. This, of course, means that to someone on \textit{terra firma}, with no applied
effort, it continues to move at a uniform speed of translation. This speed, depending upon the limit of available winds, is
arbitrary, as is its direction. This, of course, is the principle of inertia. (Galileo was so sure of this, that he had ongoing
wagers with sailors about the outcome of dropping a heavy cannonball from a high mast of a smoothly moving ship. While they all knew
if they did so while the ship remained docked at port that the ball fell down parallel to the mast, they uniformly were just as sure
that under sail the ball would fall back and strike the mast before hitting the deck. It is recorded - and presumedly not apocryphal
- that this bet was once taken, performed, and agreed with Galileo's end of the wager, that just as at dock, it fell parallel to the
mast, and hit the deck the same distance in front of the mast as it was when dropped from on high.)\\

In the second of his ``Two New Sciences", that of ``Dynamics", Galileo presents an extrapolation of his experimentation on accelerated motion,
using the then current ``winning" argument that a \textit{perpetuum mobile} is impossible, to again almost prove that a moving object,
undisturbed, continues to move at a constant velocity. This is Galileo's second establishment of the principle of inertia. It amusingly
buttresses the first. In the experiments on accelerated motion, Galileo had to defeat friction. He did so by etching a groove into his inclined
plane along which the ball moved, and then further still diminished friction by waxing the groove. Inertia manifests itself only under the
imagined limit of this friction vanishing. In an almost perfectly complimentary way, the residual friction of the smooth round ball on the small
table within the confines of the ship at full sail only \textit{enhances} the result that the ball keeps moving uniformly. (Later we shall
comment on what Galileo had to learn and contemplate about time to accomplish his novel work on accelerated motion.)\\

The theory of relativity rests upon ``relativity" - that relative motions exhaust the meaning of motion, with no recourse upon any ``absolute"
determination of motion (i.e. anti-Newtonian), \textit {and} a full import of homogeneity and isotropy of space. The extent to which isotropy is
critical is beyond any of Galileo's written thoughts. Here, we are simply nurturing the seeds he laid down.\\

The goal of this paper is to clean out, to the core, the actual underpinnings of the theories of relativity.  This is an
epistemological exercise, and so of less than high modern interest. This paper is also, to my mind, pedagogical. However it is for a
more elite audience than other such offerings intended for neophytes.  That relativity can be traced to just the ``$1^{st}$
\textit{postulate}" of Einstein has a devoted large prior body of literature:  For a cut through it see references
\cite{1}-\cite{10}.  These are all offered in 1-D where isotropy is no more than the discrete symmetry of parity.  This is
inappropriate, because both the results and the arguments they rely upon, are not sustainable in higher dimensions without some
modifications. This literature, over the decades of its existence, has led to an impression amongst ``experts" that it is a choice
whether one accepts ``isotropy" or, rather, ``the constant speed of light" as the member of a necessary set of
``\textit{postulates}". It is a purpose of this paper to carefully determine that there is no choice to be had.  (See Section 7 for a
full technical discussion.)
\\

Starting with Einstein, these ``derivations", with their so-called ``postulates", have never had the full clarity to be deemed
logical.  This is not quite Einstein's tack, where he first verifies that a constant speed of light is, at least, not inconsistent
with relativity.  Nevertheless, early on, one must ask \textit{what} symmetry justifies ``...from reasons of symmetry we are entitled
to assume that the motion of \textit{k} may be such that the axes of the moving system are at the time \textit{t} ... parallel to the
axes of the stationary system."  The reader who thinks only ``isotropy" is mistaken.  As the reader of this paper will see, even this
not only requires significant input from isotropy, but also, just as critically, from ``homogeneity", \textit{i.e.}, that the
relations will end up linear.  The careful reader should have wondered what it even
means that the axes are parallel.  (See the discussion of parallelism following (\ref{1}) in what follows.)\\

In this paper, not only do I show that the constant speed of light is unnecessary for the construction of the theories of relativity,
but overwhelmingly more, there is no \textit{room} for it in the theory.  The idea that you can choose your postulational
underpinning arises in these discussions because just one consequence of a general mathematical package is explicitly stated.  This
almost always means the rest of the package is entangled, but is left unstated, although by self-consistency, the final result has
already become determined. Instead, the argument injects other, seemingly unrelated ingredients to reach its goal.  One is using less
than one has implicitly assumed. So-called \textit{Gedankenexperiments} always illuminate this. One adds nothing fundamentally new by
such mental considerations.  Quite to the contrary, by ingenious thought, these unrecognized, but existent, tacit relations suddenly
become tangible to mind.  This is the process of how we become aware.  Nothing has been
added:  The unrealized simply mounts to high awareness, and legitimately leads the path to its correct ends, while along the course, enlightening us. \\

There are two problems about 1-D treatments.  The first is how to get to 3-D when the 1-D discussion is complete, and the second is how one ever
got there in the first place.  For example, the projective denominator of the velocity ``addition" law contains $Vv$.  To reach to 3-D, now
accepting isotropy, one says it is to be $\textbf{V}\cdot\textbf{v}$ to be a rotational invariant.  But why not $|\textbf{V}||\textbf{v}|$?  But,
remember, this $V$ was supposed to be odd under parity, although $V\rightarrow\ -V$ is \textit{also} the method of inverting the transformation,
and there is no physical role whatever for parity. Had one assumed isotropy in 3-D in the first place, one would have realized what was at stake,
and this
final, questionable massaging obviated.\\

Much more importantly is how one ever alit into what must be a 3-D geometry with a polar (\textit{not} planar) 1-D symmetry, without
ever noticing the difference between the two. The ``reasons of symmetry" comments are much too facile, because upon inspection they
already determine the outcome, and so stand as the actual but hidden postulates, the rest of
the offering all beside the point, and, fortunately not inconsistent, upon after-the-fact interpretation.  These two problems are fully dispatched here.\\

% By a degree of sloppiness of what is being assumed, together with working in environments that appear to be sufficient, but have been
%logically disingenuously entered into, one seems to get more than one has \underline{by the stated arguments}.  A space, replete with
%\textit{all} the relationships of the rotation group, is already assumed to even discuss the spherical shape of a light wave.  With that implicit
%assumption, relativity alone already determines a constant speed.  It is postulationaly redundant, and had it been otherwise,
%incompatible.\\

At the same time, this paper takes a ``high road".  We determine, at the onset, that careful thinking of isotropy forces us into the
consideration of certain rotations, related to the ``Wigner rotations", which, it turns out, are of identical form to them:  It is
impossible to correctly frame even the posit of relativity without their potential presence.  (A mathematically inclined reader may
view the greater bulk of this paper to be a lengthy constructive proof of the proposition that if this rotation is always the
identity then the theory is of Galilean-Newtonian invariance, while should it ever be non-trivial for any value of its arguments,
then the theory is of Lorentz invariance.  The thoughtful reader will realize that although these rotations are barely utilized in
the construction, their potential presence steadily guides the development.) We might have hoped that this was unnecessary. But upon
reflection, no
such rich structure could have emerged otherwise.\\

Finally, by way of introduction, I here discuss a relationship \textbf{r}(\textbf{v},\textbf{V}), (\ref{1}), which stands for a
special case of the relative velocity of the two arguments.  This is what one (starting with Einstein) calls the ``addition of
velocities", save that here it is the ``subtraction".  This ``addition" relationship, is here called \textbf{s}.  No such object or
relationship exists that \textit{mathematically can be called addition.} It does so only in certain abelian subgroups of the Lorentz
group, namely, when all the velocities are parallel.  This counts as one of the deep objections to discussions offered in just one
dimension.  Instead \textbf{s} possesses a non-commutative relationship (\ref{acom}) that should inform the unwary that things are
very different, even for the rudimentary enunciations of the principle of relativity, from what he had imagined.  A knowledgable
reader might now consider that ``addition" is, in generality, actually a non-commutative group operation. But this too is incorrect.
As we shall see, (\ref{assoc}), ``addition" is, in fact, generally also non-\textit{associative}. Such a relationship, evidently, is
nothing like addition.  It isn't a group multiplication property, because this ``addition" is a structure induced under the
projection of space-time to $\textbf{R}^{3}$, and so possesses different, projective properties. (See the discussion surrounding
(\ref{g1}).) This asymmetry between \textbf{v} and \textbf{V} shows itself most dramatically in the discussion surrounding (\ref{r}),
where we notice that no world can be constructed at $V=c$, whereas entities moving at $v=c$ are non-problematic, with $c$ emergent as
a special and ``limiting" speed.\\

%This means that not only is the 1-D ``addition" misleading, but actually incorrect.
%%%%%%%%%%%%%%%%%%%%%%%%%%%%%%%%%%%%%%%%%%%%%%%%%%%%%%%%

\section{Galilean Relativity}

When we say a motion is one of uniform translation, we mean that compared to uniformly laid out markers in a space equipped with some
uniform way of measuring time, the motion covers distance in proportion to time.  That is, we possess a physically metrized Euclidean
3-space, with coordinates \textbf{x} and a corresponding time, $t$, against which all our observations are to be compared.  An
equivalent inertial world, according to Galileo, is measured against this metrical backdrop to be comprised precisely of parts all
moving in parallel at a speed $V$, which is then the \textit{relative} velocity of it to me.  To be equivalent means first, that it
too has some similar metrization of space and time that allows it too to determine, in particular, uniform motions.  This means that
in this translating world, all the parts are to be, forever, at mutual rest.  Galileo's critical point about inertia, is that it
requests no additional effort, whatsoever, to keep them mutually at rest.  It is only such an assembly of pieces all mutually at rest
that can be rigidified into rulers, clocks, and so forth with an arbitrarily small exertion of energy.  One need not engage in this
rigidification to discuss the ensemble, merely imagining that it may be done should the collection want to measure things.  (It is
beyond Galileo, but, with foreknowledge of the development, useful to note that it may be that some parallel bundles of motions all
at speed $V$ may, somehow, fail to be at rest to one another.  Specifically, we say there is a speed $c$, such that so long as $V<c$
they are all at mutual rest, and allow the metrization of a world.  Any such allowed ensemble is a `Galilean world'.  The inequality
is strict, and adds no restriction to Galileo's thoughts should $c=\infty$.  But should
$c$ be finite then motions at $V=c$ cannot serve as the foundation of an equivalent world.)\\

Next, and most importantly, should this other equivalent world measure things, I demand to know how to compute from my observations
what is observed there.  This, of course, is the entire question of this theoretical pursuit.  In particular I demand to know its
determination of uniform motions.  To fully do so, I must establish the relation of its metrization of space and time compared to
mine.  This is the goal of the theory.\\

Galileo accomplished such a theory by his assumption of the `principle of relativity', with `relative' here the antithesis of
`absolute'. Specifically, and in mathematical implementation, the principle states that when an equivalent inertial world measures
another at velocity \textbf{V} relative to itself, then the rule it uses to compute the new coordinates in the equivalent
\textbf{V}-world are precisely the same as I (any other equivalent world) use to compute the coordinates of a world at relative
velocity \textbf{V} to myself. Since this applies only to these allowed inertial worlds, and is constructed through considerations of
uniform (inertial) motions at allowed speeds, I shall more precisely refer to this as the `principle of inertial relativity'. This is
Galileo's principle utilized in his thoughts.  This precision of language is intended to be employed when we discriminate against an
attempt to erect such a theory by \textit{never} invoking inertial relativity, but rather doing so on appeal only to these potential
$c$
motions.\\

We start by assuming, with Galileo, that all uniform translating motions in one world (``frame") are also (other) uniform translating
motions in another. (This is necessary for `inertial equivalence' to be an equivalence relation.) More sharply, we mean that all
trajectories of a common fixed velocity (vector) in one world are also trajectories of (generally another) common fixed velocity in
any other. This means we can discuss relations between velocities independent of where or when. When we say ``where or when", we are
further informing the reader that velocities are indeed a coordination between some underlying space and time. Moreover, we intend,
following Galileo and ``common sense" to understand that when two moving entities are seen to meet at some point at some instant of
time, then they also appear to do so in any other world. This means, prior to velocity relations, we have assumed a space-time point
relationship existing between worlds. We shall exhaust the treatment of the
circumstances and consequences of these assumptions of ``homogeneity" in the next section, but for now, discuss properties of velocity.\\

I sit in a Galilean world. I can measure the speeds of things everywhere. This means I am replete with rulers, clocks and so forth over a vast
expanse of space. I shall call this ``world" ``I" (either myself as ``I" or roman numeral world I). I can watch another equally equipped world
(``frame") moving at velocity \textbf{V} I can measure. By this I mean that every particle it is constructed out of, I measure to have speed
\textbf{V}. I shall also denote this world as a
world by the symbol V.\\

Not only do I see the world V, but I also see various particles moving at various velocities \textbf{v}. The world V also sees these particles at
some other velocity, and I reserve the name \textbf{r} for a determined rule that allows me to calculate, given V's velocity \textbf{V} that I
measure and a particle's velocity \textbf{v} that I measure, the \textit{relative} velocity with which V sees the particle to be moving at:

\begin{equation}
\textbf{r}(\textbf{v},\textbf{V}). \label{1}
\end{equation}

There are two critical comments to make. First the order of the arguments matters: \textbf{v} is the velocity of some particle, but \textbf{V} is
that of a full-blown world that can measure things everywhere.  The velocity of a world can appear as the first argument, by taking \textbf{v} to
be that of any of the particles that constitute that world, and just so \textbf{r} can be construed as the velocity of a world.  The second
argument, however, must be the velocity of a world, which possesses the requisite structure to determine directions, and so forth:  \textbf{V} is
a \textit{parameter} determining the rule, and \textbf{v} an independent variable that rule is applied to.  In what follows, upper or lower case
must never, on its own, determine the meaning of the symbol:  Only the position of the argument does, so that if a particle's velocity becomes
promoted to the second argument, the result has meaning only when it is understood that it now is the velocity of a constituent of a world at
that velocity.
\\

Secondly, I immediately have to address isotropy. My world is isotropic. So is V's. This means I and V, quite independently, can
\textit{arbitrarily} assign axes with respect to which directions are measured with complete impunity. So, my declaration of
\textbf{r}, to have any directional meaning, requires some posit. I posit (\ref{1}) to be the rule when I have copied V in having all
our axes ``parallel". This
makes rule (\ref{1}) definite, and as shall emerge, is a highly non-trivial detail.\\

What I mean by ``parallel" is the following:  I have arbitrarily erected some reference $X,Y,Z$-axes, in terms of which I see
particles of V streaming in the direction of $\widehat{\textbf{V}}$.  My full isotropic ignorance has been broken by this referent,
and I can conveniently align my new $x$-axis in this direction. A line of particles of V now forever moves along this axis. But then
V can choose this same line as his new $x$-axis, and so sees my particles streaming in the direction of $-\widehat{\textbf{x}}$,
although at this point, at undetermined \textit{speed}. V next chooses an $(x,y)$-plane containing his $x$-axis and an (arbitrarily
chosen) orthogonal $y$-axis, and illuminates lines for various positive values of $y$=constant in it. But I must see these as
coplanar as well:  These are lines of particles all in the direction of $\widehat{\textbf{x}}$, and so constitute a (cylindric) ruled
surface for me.  Any particle moving uniformly in V's $(x,y)$-plane transverse to the lines $y$=constant must appear to me on this
ruled surface. But, uniform motions to V are uniform motions to me (\textit{i.e.} straight lines), and so the ruled surface must be a
plane. I now choose my $y$-axis, in this plane, orthogonal to my $x$-axis, and in the direction that I see V's illuminated lines.
Finally, we both agree to use the right hand rule to determine our $z$-axes. V and I are now aligned. We are identical vector spaces,
and can agree to rotate, the both of us identically, to some other chosen orientation, should we choose to do so. For example, I
revert to my original $X,Y,Z$-axes, by a rotation determined from the direction cosines of my $x,y,z$-axes, and then inform V to
perform precisely this rotation on his $x,y,z$-axes, so that, in particular, I see V at $\widehat{\textbf{V}}$, and V sees me at
-$\widehat{\textbf{V}}$. This then is what it means to be ``parallel". (While the direction $\widehat{\textbf{V}}$ is evidently
parallel in the two worlds, and orientations agree, we have not said that the planes perpendicular to $\widehat{\textbf{V}}$ in the
two worlds are actually parallel at what moment to whom.  Our procedure suffices for this Section.
)\\

Notice, \textit{inter alia}, that all particles moving in V's $(x,y)$-plane are also in I's $(x,y)$-plane, so that in this case
\textbf{v}, \textbf{V}, and $\textbf{r}(\textbf{v},\textbf{V})$ are coplanar under parallelism.  But the plane containing
$\widehat{\textbf{V}}$ that V has chosen is arbitrary, so that it follows that \textbf{v}, \textbf{V}, and
$\textbf{r}(\textbf{v},\textbf{V})$ are always coplanar under parallelism. (This is the content for \textbf{r} of polar isotropy
about $\widehat{\textbf{V}}$.  Notice, under isotropy, it is the set of polar planes, \textit{i.e.} those containing
$\widehat{\textbf{V}}$, that are parallel, not, necessarily, the planes orthogonal to it. No discussion of when is necessary for
these polar planes, since they remain fixed (and parallel) independent of time.)  The following three bracketed paragraphs should be
skipped for continuity of development, but might assist the uncomfortable reader.
\\

[Let me interrupt the development here, and offer the reader some critically useful comments.  All the paragraphs above, however
verbal they may seem to be, are pithy mathematical statements.  When I say isotropy, I mean everything implied by invariance under
the rotation group $\textbf{SO}_3$.  This leads, specifically, when I say \textbf{r} enjoys isotropy, to (\ref{23n}).  For
pedagogical assistance, I shall remind the reader of what is intended.  There is some definite three dimensional space filled up with
various fields of parallel velocities.  Some are at \textbf{v} of \textbf{0} and another set at \textbf{v} of \textbf{V}.  The first
I am calling the world I, and the second, the world V.  (This is Galileo's definition of a world.  Einstein of 1905 requests a world
to be infinitely rigid.  Einstein's might entail an ellipsis that all these inertial particles might now be easily rigidified.  Only
Galileo's definition is correct.) Physically, an observer can be imagined to view this from a distant void, so that its behavior
cannot depend upon how that observer rotates himself. Without distant referents, this is equivalent to saying that the physics of
these moving particles cannot be changed when each is identically rotated. From the definition of the collections I and V, the
distant observer's rotation is equivalent to the identical rotation of each, which then can change nothing of their physical
relation, whence the comment about the
identical rotations which maintain the parallelism of the worlds two paragraphs ago.\\

Next, when I say \textbf{r} enjoys isotropy, I mean the following deduction (the `boiler plate' of symmetry).  The distant observer
rotates by $\textbf{R}^{-1}$, and so now sees what had been \textbf{v}, \textbf{V}, and \textbf{r}(\textbf{v},\textbf{V}) as
$\textbf{R}\cdot \textbf{v}$, $\textbf{R}\cdot \textbf{V}$, and $\textbf{R}\cdot \textbf{r}(\textbf{v},\textbf{V})$.  But to say the
rule (mapping) \textbf{r} enjoys isotropy is to say that \textbf{r} performs its same task, literally unchanged, when applied in this
equivalent, rotated world.  That is, $\textbf{r}(\textbf{R}\cdot \textbf{v},\textbf{R}\cdot \textbf{V})=\textbf{R}\cdot
\textbf{r}(\textbf{v},\textbf{V})$, as shall be recalled to the reader in (\ref{23n}).  This is then the mathematical incarnation of
the physical statement of an object's rotational invariance.  The reader should pause, thinking about the worlds I and V that have
now picked up some relations between themselves. This statement is profound and powerful because it constrains how V metrizes his
space and time, and we are saying it must be in accordance with this externally observed indifference to rotation.  %(The knowledgable reader
%will understand we must do this to maintain the continued presence of the conservation of angular momentum.) This is the master
%ingredient tacit in the construction of relativity, but consciously forgotten (and out of context in 1D) in the literature. That is,
%we anticipate that V may metrize his space at an instant of his time differently from how I metrized his at an instant of his time.
%Yet, the metrical relations just of space implied by rotations are to be unaltered, which evidently must then impact upon how
%his time must consistently be remetrized. This is what the reader should have guessed he's now buying into.
\\

Now, when I say that \textbf{v}, \textbf{V}, and $\textbf{r}(\textbf{v},\textbf{V})$ are always coplanar under parallelism, I
understand a simple mathematical result following from \textbf{r}'s isotropy.  Consider the polar subgroup of rotations about the
$x$-axis (\textit{i.e.} the $\widehat{\textbf{V}}$-axis).  These transform the $(x,y)$ plane to any other polar plane, while leaving
\textbf{V} invariant. But, \textbf{r} maps the $(x,y)$ plane to itself under the convention of parallelism. Then, a polar rotation
maps the span of \textbf{r} to the same polar plane as it did the $(x,y)$ plane, which then by (\ref{23n}) determines that
\textbf{v}, \textbf{V}, and $\textbf{r}(\textbf{v},\textbf{V})$ are coplanar for \textbf{v} in any polar plane.  But then this is
true for all \textbf{v}.]
\\

I know one certain thing about rule \textbf{r}:

\begin{equation}
\textbf{r}(\textbf{V},\textbf{V})=\textbf{0}. \label{2}
\end{equation}
This simply says, the world V is built out of parts all moving at speed \textbf{V}, and so all at rest with respect to - the fabric
of - world V.\\

%(This is a substantial and deep observation of Galileo, that a parallel field of inertial motions in one frame is again a parallel
%field in another.  That every trajectory of uniform velocity to one appears as uniform to another determines space-time coordinates
%to be just projectively related to one another.  Only this stronger request of parallel fields of motion, developed in Section 3,
%determines that space-time coordinates are linearly related.  That even the most insubstantial of objects happily reposes at rest in
%the moving ship of the Introduction is Galileo's determination that a `frame' is precisely the ensemble of all and only those
%entities moving inertially at velocity \textbf{V}.  As such, a `frame' is infinitely tenuous, although parts of it can now be glued
%together with arbitrarily little expenditure of work to grant the rigidity required for measurements, etc.  Had the worlds been
%merely projectively related, the different particles moving inertially at velocity \textbf{V} would still be relatively moving to one
%another, and so their gluing together would cost an arbitrarily large exertion.)\\

With the sharp proviso of rule \textbf{r}, that our axes are parallel, it can also be concluded that

\begin{equation}
\textbf{r}(\textbf{v},\textbf{0})=\textbf{v}. \label{3}
\end{equation}

This says a world at rest with respect to me \textit{and with axes parallel to mine}, observes, just as I do, a particle at velocity \textbf{v}
to me at \textbf{v} to him. Had we allowed V(\textbf{V}=\textbf{0}) to arbitrarily rotate its axes from mine, then (\ref{3}) would be correct
only up to a rotation. The rule (\ref{1}) has
legislated that rotation to be the identity.\\

So far I have my world and the equivalent world V. I next observe a third equivalent world V$'$. V$'$ also aligns his axes with me. But now, by
(\ref{1}), V$'$ sees each particle of V, and hence V itself at velocity

\begin{equation}
\widetilde{\textbf{V}}=\textbf{r}(\textbf{V},\textbf{V}'). \label{4}
\end{equation}

V$'$ also sees the particle I see at velocity \textbf{v} at
\begin{equation}
\widetilde{\textbf{v}}=\textbf{r}(\textbf{v},\textbf{V}'). \label{4pr}
\end{equation}
But, V$'$ is an identical world to mine, and so V$'$ can determine that the world V with axes parallel to V$'$, sees the particle at velocity

\begin{equation}
\textbf{v}''=\textbf{r}(\widetilde{\textbf{v}},
\widetilde{\textbf{V}})=\textbf{r}(\textbf{r}(\textbf{v},\textbf{V}'),\textbf{r}(\textbf{V},\textbf{V}')).
\label{5}
\end{equation}

However, by (\ref{1}), I can say that V, with axes parallel to me, sees the particle at

\begin{equation}
\textbf{v}'=\textbf{r}(\textbf{v},\textbf{V}). \label{6}
\end{equation}

Since only relative motions have meaning, we are sorely tempted to equate $\textbf{v}'=\textbf{v}''$. What I may \textit{only} conclude is that

\begin{equation}
|\textbf{r}(\textbf{v},\textbf{V})|=
|\textbf{r}(\textbf{r}(\textbf{v},\textbf{V}'),\textbf{r}(\textbf{V},\textbf{V}'))|:
\label{7}
\end{equation}

Upon reflection, granted the isotropy of all these worlds, I cannot conclude the transitivity of parallelism: If V$'$ is parallel to me, and V
parallel to V$'$, I \textit{cannot} - I have no means to logically decide the issue - assert that V is then also parallel to me: It might well be
rotated. So,

\begin{equation}
\textbf{r}(\textbf{r}(\textbf{v},
\textbf{V}'),\textbf{r}(\textbf{V},\textbf{V}'))=\textbf{R}(\textbf{V}',\textbf{V})\cdot
\textbf{r}(\textbf{v},\textbf{V}) \label{8}
\end{equation}

with \textbf{R} some rotation, continuous in its dependence on \textbf{V}$'$ and \textbf{V}, since the \textbf{r}'s are. (The
possibility of \textbf{R} was never
 contemplated by Galileo. Nor was it explicitly by Lorentz, Poincar\'{e},
 nor the Einstein of 1905.) But with careful consideration, $\textbf{R}\neq
 \textbf{1}$ can't be ruled out. (It is termed a ``Wigner rotation", which rather than some piece of esotericism, is fundamental to the theory
 of relativity.)  However, we know one extra thing by
 isotropy. Our consideration is as follows.\\

When I determine how the world V$'$ sees things, the formula $\textbf{r}(\cdot,\textbf{V}')$ entails a breakdown of totally isotropic space into
a polar one, with polar axis along \textbf{V}$'$, but then with polar isotropy about that axis. Should it turn out that \textbf{r} leaves the
total isotropy intact, then \textbf{R} of (\ref{8}) would be the identity. If it actually does polarize space, then the problem of \textbf{R}
arises. This follows, since when V$'$ looks out along $\widetilde{\textbf{V}}$, he polarizes his space on yet another axis, and so a rotation
\textbf{R} is the \textit{best} to be expected. Were (\ref{8}) worse than a rotation, then any principle of relative velocities would necessarily
fail, and a radically new
conception of perceived motions would be required.\\

However, there is one easy case. Should \textbf{V} and \textbf{V}$'$ be parallel, then only one polarization is requested, so that in
a world with the maximally remaining symmetry - isotropy about the one, unique, axis - there again can be no rotation. (This is done
\textit{de facto} in most presentations by working in just one spatial dimension. This is at best disingenuous, and as transpires,
deeply misleading.) So, with attached proviso

\begin{equation}
\textbf{r}(\textbf{r}(\textbf{v},
\textbf{V}'),\textbf{r}(\textbf{V},\textbf{V}'))=\textbf{r}(\textbf{v},\textbf{V}),
\ \ \textbf{V}\|\textbf{V}', \ \ \textrm{or} \ \
\textbf{R}(V'\widehat{\textbf{V}},V\widehat{\textbf{V}})=\textbf{1}.
\label{9}
\end{equation}

Together with general isotropy, (\ref{9}) will prove to
service our needs.\\

[This reduction of \textbf{R} to the identity is a simple lemma.  \textbf{V} and \textbf{V}$'$ parallel allows that either might
vanish.  It is sufficient to consider that \textbf{V}$'$ might.  But with \textbf{V}$'$ vanishing, (\ref{8}), together with
(\ref{3}), immediately implies that \textbf{R} is the identity. Next, with \textbf{V}$'$ non-vanishing, the span of \textbf{V} and
\textbf{V}$'$ is just that of \textbf{V}$'$ whether or not \textbf{V} vanishes.  But, $\textbf{R}(\textbf{V}',\textbf{V})$, a
rotation of worlds, is independent of \textbf{v}.  With \textbf{v} equal to $\textbf{V}'$, by parallelism, we see by (\ref{8}) that
\textbf{R} leaves the span of \textbf{V}$'$ invariant, and then for for any other \textbf{v}, its span with \textbf{V}$'$ invariant.
But since it leaves \textbf{V}$'$ invariant, then it must also leave the span of \textbf{v} invariant.  (In any number of dimensions,
\textbf{R} restricted to this 2-D invariant subspace is just a 2-D rotation in that plane.  Only a 0 or $\pi$ rotation can leave the
span of one axis invariant.) Thus, under \textbf{R}, each basis vector of the space maps to either itself or its negative. But
\textbf{R} is the identity when \textbf{V}$'$ vanishes, and \textbf{R} is continuous in its arguments.  Hence each sign is plus, and
\textbf{R} is the identity.
This, indeed, is true in all spatial dimensions.] \\

Formula (\ref{9}), or its more general version (\ref{8}) is Galileo's ``Principle of Relativity", used implicitly by him (and by
Einstein in the
same vein) where required. %Here, I have strived to put all thought and assumption in full, transparent view, with, as shall transpire, critical
%consequence.
(\ref{8}) and (\ref{9}) are the expression that when all worlds are equivalent, only relative motions matter, and so strong
consistency relations inhere. \textit{This} is the principle of relativity. Let us see what
it dictates.\\

Set $\textbf{v}=\textbf{V}'$ in (\ref{8}), and use (\ref{2}):

\begin{equation}
\textbf{r}(\textbf{0},\textbf{r}(\textbf{V},
\textbf{v}))=\textbf{R}(\textbf{v},\textbf{V})\cdot
\textbf{r}(\textbf{v},\textbf{V}). \label{10}
\end{equation}

Next, set $\textbf{V}=\textbf{0}$ in (\ref{10}), use (\ref{3}), and notice that since $\textbf{0}\|\textbf{v}$, $\textbf{R}=\textbf{1}$:

\begin{equation}
\textbf{r}(\textbf{0},\textbf{r}(\textbf{0},\textbf{V}))=\textbf{V},
\label{11}
\end{equation}

renaming $\textbf{v}\rightarrow\textbf{V}$. Next, denote

\begin{equation}
\textbf{b}(\textbf{V})\equiv\textbf{r}(\textbf{0},\textbf{V})
\label{12}
\end{equation}

so that (\ref{11}) reads

\begin{equation}
\textbf{b}(\textbf{b}(\textbf{V}))=\textbf{V}, \textbf{b}(\textbf{0})=\textbf{0} \ \textrm{by (\ref{12}) and (\ref{3}).} \label{13}
\end{equation}

(\ref{12}) defines \textbf{b} to be the velocity with which V sees me moving when I see him at \textbf{V}. By (\ref{13}),

\begin{equation}
\textrm{if} \ \textbf{b}(\textbf{V})=\textbf{0} \ \textrm{then} \
\textbf{V}=\textbf{b}(\textbf{0})=\textbf{0} \ : \
\textbf{b}(\textbf{V})\neq{\textbf{0}} \ \textrm{if} \
\textbf{V}\neq{\textbf{0}},
 \label{14}
\end{equation}

so that \textbf{b} is invertible about \textbf{V}=\textbf{0}. But, V assigns me the direction of $-\widehat{\textbf{V}}$, by definition of our
parallelism. So,

\begin{equation}
\textbf{b}(\textbf{V})=-\widehat{\textbf{V}}u(|\textbf{V}|).
\label{15}
\end{equation}

Entering (\ref{15}) in (\ref{13}), we have

\begin{equation}
\widehat{\textbf{V}}u(u(|\textbf{V}|))= \widehat{\textbf{V}}|\textbf{V}|, \ \textrm{or} \ u(u(|\textbf{V}|))=|\textbf{V}|,u(0)=0.
\label{16}
\end{equation}

Since $u(u(x))=x$, should $u(x)=u(y)$, applying $u$ to both sides yields $x=y$, so that $x\neq{y}$ implies $u(x)\neq{u(y)}$. So, if
$u$ is continuous then it is strictly monotone.  But $u$ maps the positive $x$-axis to the positive $x$-axis and vanishes at $x=0$.
For any $x$ other than $0$ in its domain it is non-negative, and so it must be strictly monotone increasing. Consider that $u(x)>x$.
Then, $u(u(x))>u(x)$ and so, $x>u(x)$.  This is impossible, as is similarly the case $u(x)<x$.  It now follows that $u(x)=x$, and so

\begin{equation}
\textbf{r}(0,\textbf{V})=-\textbf{V}. \label{17}
\end{equation}

There is no proviso to (\ref{17}): The rule \textbf{r} assumes parallel axes, and (\ref{17}) then says that if I see V at \textbf{V}, then V sees
me at -\textbf{V}.
 However intuitive, the deduction of (\ref{17}) is non-trivial, as then
 too must be its content. In particular, using (\ref{17}) in (\ref{10}), we
 have

 \begin{equation}
-\textbf{r}(\textbf{V},\textbf{v})=\textbf{R}(\textbf{v},\textbf{V})\cdot
\textbf{r}(\textbf{v},\textbf{V}). \label{18}
\end{equation}

So

\begin{equation}
|\textbf{r}(\textbf{V},\textbf{v})|=|\textbf{r}(\textbf{v},\textbf{V})|,
\label{19}
\end{equation}

and the relative \textit{speed} of two particles is independent of which we consider. Should I notice them to be \textit{parallel}, then

\begin{equation}
-\textbf{r}(\textbf{V},\textbf{v})=\textbf{r}(\textbf{v},\textbf{V})
\ \textbf{v}\|\textbf{V}. \label{20}
\end{equation}

\noindent However obvious (\ref{20}) might appear to be, bear in mind that when I see \textbf{v} and \textbf{V} non-parallel, then (\ref{20})
is quite apt to be false, correct only up to a rotation, (\ref{18}). \\

The 1905 argument of Einstein - and his later pedagogic offerings - always offered two ``postulates" of the special theory of relativity. The
first is Galileo: All frames in relative uniform translation are equivalent. The second is ``The speed of light is identical in all such frames."
Together with all subsequent treatments, it is a misuse of language to call these \textit{the} ``postulates". We notice in all of these offerings
hand-wavings about linearity of equations, the first ``postulate" meaning that V $\rightarrow$ -V is an inverse, and so forth. Again, here every
relied-upon detail of deduction shall have been laid out. \\

The power of (\ref{20}), as we shall soon learn, is that it \textit{totally} supplants the second ``postulate" about the universal speed of
light. The reader will learn this shortly. (\ref{20}) is not obvious, but of exceeding potency. There is no role for light in the special and
general theories of relativity of
Einstein.\\

We next demonstrate that the relation \textbf{r} is always invertible, and indeed quite simply. Substitute \textbf{V}=\textbf{0} in (\ref{8}),
use (\ref{17}), (\ref{3}), and then \textbf{R}(\textbf{V}$'$,\textbf{0})=\textbf{1}, since V$'\|$\textbf{0}, and finally rename
\textbf{V}$'\rightarrow$ \textbf{V}:

\begin{equation}
\textbf{r}(\textbf{r}(\textbf{v},\textbf{V}),-\textbf{V})=\textbf{v}.
\label{21}
\end{equation}

That is,
\begin{equation}
\textbf{r}(\textbf{v},\textbf{V})=\textbf{u} \Rightarrow
\textbf{v}=\textbf{r}(\textbf{u},-\textbf{V}). \label{22}
\end{equation}

Formulas (\ref{17}) and (\ref{22}) establish \textit{for velocities}, the substitution \textbf{V} $\rightarrow$ -\textbf{V} is precisely the
inverse transformation from the world V back to me. Together with (\ref{20}) in homogeneous isotropic worlds, this \textit{completely} determines
the special theory of relativity. \\

Let us continue this section on relativity with an interlude about the rotation \textbf{R} of (\ref{8}).  Repeating (\ref{8}) for ease of
reference,
$$\textbf{r}(\textbf{r}(\textbf{v}, \textbf{V}'),\textbf{r}(\textbf{V},\textbf{V}'))=\textbf{R}(\textbf{V}',\textbf{V})\cdot
\textbf{r}(\textbf{v},\textbf{V}),$$ consider any \textbf{v} in the (\textbf{V},\textbf{V}$'$)-plane.  Then by parallelism \textbf{r}(\textbf{v},
\textbf{V}$'$) is also in this plane. Similarly, \textbf{r}(\textbf{v}, \textbf{V}) is again in this plane, as certainly too is
\textbf{r}(\textbf{V},\textbf{V}$'$).  From the first and third of these observations, we now conclude that the entire LHS is also in the
(\textbf{V},\textbf{V}$'$)-plane.  But then \textbf{R}(\textbf{V}$'$,\textbf{V}) is a rotation that leaves the (\textbf{V},\textbf{V}$'$)-plane
invariant, and hence is a rotation about the axis $\textbf{V}'\wedge\textbf{V}$. \\

Recall in the discussion of parallelism, that both V and I are identical vector spaces, and so relationships, and in particular \textbf{r} of
(\ref{1}), are preserved upon the identical rotation of the two worlds.  This means \textbf{r} is a rotational vector, so that for \textit{any}
rotation \textbf{R}, the application of the same rule \textbf{r} to its rotated arguments is that same rotation on \textbf{r} of its original
arguments:

\begin{equation}
\textbf{R}\cdot \textbf{r}(\textbf{v},\textbf{V})=\textbf{r}(\textbf{R}\cdot \textbf{v},\textbf{R}\cdot \textbf{V}). \label{23n}
\end{equation}

Consider (\ref{8}), with different variables,
$$\textbf{r}(\textbf{r}(\widetilde{\textbf{v}},\widetilde{\textbf{U}}),\textbf{r}(\widetilde{\textbf{V}},\widetilde{\textbf{U}})
)=\textbf{R}(\widetilde{\textbf{U}},\widetilde{\textbf{V}})\textbf{r}(\widetilde{\textbf{v}},\widetilde{\textbf{V}})$$ where each of the tilde'd
objects is of the form  $$\widetilde{\textbf{X}}=\textbf{r}(\textbf{X},\textbf{V}').$$ By (\ref{8}), the LHS is
$$\textbf{r}(\textbf{R}(\textbf{V}',\textbf{U})\cdot
\textbf{r}(\textbf{v},\textbf{U}),\textbf{R}(\textbf{V}',\textbf{U})\cdot \textbf{r}(\textbf{V},\textbf{U})),$$ which by (\ref{23n}), is
$$\textbf{R}(\textbf{V}',\textbf{U})\cdot \textbf{r}(\textbf{r}(\textbf{v},\textbf{U}),
\textbf{r}(\textbf{V},\textbf{U})),$$ which, by another application of (\ref{8}), is
$$\textbf{R}(\textbf{V}',\textbf{U})\cdot \textbf{R}(\textbf{U},\textbf{V})\cdot\textbf{r}(\textbf{v},\textbf{V}).$$
The RHS is, again by(\ref{8}),
$$\textbf{R}(\widetilde{\textbf{U}},\widetilde{\textbf{V}})\cdot\textbf{R}(\textbf{V}',\textbf{V})\cdot
\textbf{r}(\textbf{v},\textbf{V}).$$ Equating the above two lines, with \textbf{v} arbitrary, and \textbf{r} invertible, we conclude
\begin{equation}
\textbf{R}(\textbf{V}',\textbf{U})\cdot
\textbf{R}(\textbf{U},\textbf{V})=\textbf{R}(\textbf{r}(\textbf{U},\textbf{V}'),\textbf{r}(\textbf{V},\textbf{V}'))
\cdot\textbf{R}(\textbf{V}',\textbf{V}). \label{R2}
\end{equation}
Equation (\ref{R2}) leads to two powerful conclusions.\\

First set $\textbf{V}'=\textbf{V}$, use (\ref{2}), and two applications of (\ref{9}), and determine that
$$\textbf{R}(\textbf{V},\textbf{U}) \cdot \textbf{R}(\textbf{U},\textbf{V})=\textbf{1},$$

so that
\begin{equation}
\textbf{R}(\textbf{V}',\textbf{V})^{-1}=\textbf{R}(\textbf{V},\textbf{V}'). \label{24n}
\end{equation}

Together with the proviso of isotropy of (\ref{9}), that $\textbf{R}(V'\widehat{\textbf{V}},V\widehat{\textbf{V}})=\textbf{1}$, (\ref{24n})
fundamentally restricts \textbf{R} to be a rotation about axis $\textbf{V}'\wedge\textbf{V}$, with axis and angle of rotation of the form
$$\hat{\textbf{n}} \sin(\theta)=\textbf{V}' \wedge \textbf{V} \; s(\textbf{V},\textbf{V}')$$
with $s$ a rotational scalar symmetric in its arguments.  It's explicit form will appear later, (\ref{69}). \\

As a second application of (\ref{R2}), set $\textbf{U}=\textbf{0}$.  It is easy to see, that with (\ref{24n}),
\begin{equation}
\textbf{R}(\textbf{V},\textbf{V}')=\textbf{R}(-\textbf{V}',\textbf{r}(\textbf{V},\textbf{V}')). \label{wig}
\end{equation}
We offer this, for now, as an extra, strong, structural relationship on \textbf{R}.  Later (see the discussion following (\ref{g1})), we shall
notice that its RHS is \textit{precisely} a ``Wigner rotation", so that, up to some sign convention, so too is \textbf{R}.  It is striking, the
extent to which its inner relations have been
determined purely in consequence of relativity.\\

We conclude this section with the mathematical properties of the so-called ``addition of velocities".  This object, with \textbf{s} for ``sum",
is
\begin{equation}
\textbf{s}(\textbf{V},\textbf{V}')\equiv \textbf{r}(\textbf{V},-\textbf{V}'). \label{add}
\end{equation}
First of all, the sharp distinction of the order of the arguments holds exactly as stated in the comment after (\ref{1}).  Next, we can restate
(\ref{18}), using a trivial version of (\ref{23n}) with an appropriate $\pi$ rotation as $-
\textbf{r}(\textbf{v},\textbf{V})=\textbf{r}(-\textbf{v},-\textbf{V})$, as
\begin{equation}
\textbf{s}(\textbf{V},\textbf{V}')= \textbf{R}(\textbf{V}',-\textbf{V})\cdot \textbf{s}(\textbf{V}',\textbf{V}), \label{acom}
\end{equation}
so that, unless \textbf{V} and $\textbf{V}'$ are parallel, \textbf{s} is non-commutative.\\

Again write down (\ref{8}), with, for lexical convenience, these different variables:
$$\textbf{r}(\textbf{r}(\textbf{v},
\textbf{V}),\textbf{r}(-\textbf{W},\textbf{V}))=\textbf{R}(\textbf{V},-\textbf{W})\cdot \textbf{r}(\textbf{v},-\textbf{W}).$$ Next, set
$\textbf{U}=\textbf{r}(\textbf{v}, \textbf{V})$, use (\ref{22}) to eliminate \textbf{v} in terms of \textbf{U}, and use (\ref{24n}) to place the
\textbf{R} on the LHS to obtain
$$\textbf{R}(-\textbf{W},\textbf{V})\cdot \textbf{r}(\textbf{U},\textbf{r}(-\textbf{W},\textbf{V}))
=\textbf{r}(\textbf{r}(\textbf{U},-\textbf{V}),-\textbf{W}).$$ Next move \textbf{R} inside \textbf{r} by (\ref{23n}), and use (\ref{18}) to
obtain
$$\textbf{r}(\textbf{R}(-\textbf{W},\textbf{V})\cdot \textbf{U},-\textbf{r}(\textbf{V},-\textbf{W}))
=\textbf{r}(\textbf{r}(\textbf{U},-\textbf{V}),-\textbf{W}),$$ which, in terms of \textbf{s}, by its definition, (\ref{add}), becomes
\begin{equation}
\textbf{s}(\textbf{R}(-\textbf{W},\textbf{V})\cdot \textbf{U},\textbf{s}(\textbf{V},\textbf{W}))
=\textbf{s}(\textbf{s}(\textbf{U},\textbf{V}),\textbf{W}). \label{assoc}
\end{equation}
Equation (\ref{assoc}) specifies how ``addition" is \textit{non}-associative.  To be precise, with \textbf{r}, and hence \textbf{s}, invertible
on its first argument, \textbf{s} is associative if, and only if, $\textbf{R}(-\textbf{W},\textbf{V})\cdot \textbf{U}=\textbf{U}$, which is to
say that either \textbf{U} is along the axis of the rotation \textbf{R} (that is, orthogonal to the span of \textbf{V} and \textbf{W}), or for
\textit{any} \textbf{U}, if, and only if, \textbf{V} and \textbf{W} are parallel. Equations (\ref{acom}) and (\ref{assoc}) spell out the
elementary properties of \textbf{s}, which evidently, has no kinship to a relation termed ``addition".
%%%%%%%%%%%%%%%%%%%%%%%%%%%%%%%%%%%%%%%%%%%%%%%%%%%%%%%%

\section{Galilean Worlds Are Linearly Related}

It is easy to show that Galileo's worlds, with parallel uniform velocities in one appearing as parallel uniform velocities in another, must be
linearly related. More precisely, if the spatial coordinates and time of one world are related to those of another by a 1:1 point transformation

\begin{equation}
\textbf{x}'=\textbf{f}(\textbf{x},t),\ \ t'=g(\textbf{x},t),
\label{23}
\end{equation}

then necessarily both \textbf{f} and $g$ are linear in their
arguments. \\

Galileo satisfied himself with what turns out to be the simplest candidate for a world \textit{system} (\textit{i.e.} the relations between all
the worlds at different uniform translations), with

\begin{equation}
\textbf{x}'=\textbf{x}-\textbf{V}t,\ \ t'=t. \label{24}
\end{equation}

(\ref{24}) possesses \textit{two} deep conceptual simplifications upon \textit{any} system that meets all of his demands. Not only
does the form (\ref{24}) \textit{not} polarize space, and so free of the conceptually difficult rotations \textbf{R} that make mutual
relative velocities more complicated than one might have intuited (II.\ref{18}), but further has these worlds all sharing a universal
time. I believe the rotations (II.\ref{10}) and (II.\ref{18}) far exceeded Galileo's conceptual world. In the next section, I shall
discuss Galileo's conception (if not invention) of time and express that he could have intuited (\ref{23}). This next section is
mandatory here, since synchronization and so forth must be demonstrated to have neither logical nor experimental connection to
the properties and existence of light.\\

Let us pose and prove the proposition of this section.
\\

$\textrm{If for all uniform motions } $

\begin{equation}  \textbf{x}=\textbf{v}t+ \xi , \; i.e. \textrm{ for all} \;
\textrm{v} \; \textrm{and} \; \xi \; \Rightarrow \; \textbf{x}'=\textbf{r}(\textbf{v})t'+ \textbf{a}(\xi,\textbf{v})
\;\\
\textrm{, then (\ref{23}) is linear.} \label{25}
\end{equation}

\underline{Proof}: Substitute (\ref{23}) in the second of (\ref{25}), set $\textbf{x}=\textbf{v}t+ \xi$, and so,

\begin{equation}
\textbf{f}(\xi +\textbf{v}t,t)=\textbf{r}(\textbf{v})g(\xi
+\textbf{v}t,t)+ \textbf{a}(\xi,\textbf{v}) \;  \textrm{all} \;
\xi,\textbf{v},\textbf{t}. \label{26}
\end{equation}
In order for $\textbf{v}'=d\textbf{x}'/dt'$ to exist, both \textbf{f} and $g$ must be differentiable on all their arguments.  By (\ref{26}), this
implies that $\textbf{a}(\xi,\textbf{v})$ is differentiable on its first vector argument.\\

Since (\ref{26}) is correct for any $\xi$, set $\xi= \textbf{x}-\textbf{v}t$, and so,

\begin{equation}
\textbf{f}(\textbf{x},t)=\textbf{r}(\textbf{v})g(\textbf{x},t)+
\textbf{a}(\textbf{x}- \textbf{v}t,\textbf{v}). \label{27}
\end{equation}

Evaluate (\ref{27}) at \textbf{v}=\textbf{0}:

\begin{equation}
\textbf{f}(\textbf{x},t)= \hat{\textbf{r}} g(\textbf{x},t)+
\textbf{a}(\textbf{x},\textbf{0}) \; ; \; \hat{\textbf{r}} \equiv
\textbf{r}(\textbf{0}). \label{28}
\end{equation}

Differentiating on $t$, this is $\textbf{f}_{t}=\hat{\textbf{r}}g_{t}$.  With $\hat{\textbf{r}}\neq \textbf{0}$, if $g_{t}=0$, then
so too must $\textbf{f}_{t}$.  But this simultaneous vanishing makes the Jacobian singular.  Thus, $g_{t}(\textbf{x},t)\neq 0$ for
any value of its arguments.  If $\hat{\textbf{r}}= \textbf{0}$, then $\textbf{f}_{t}= \textbf{0}$, and so, again
$g_{t}(\textbf{x},t)\neq 0$ for the Jacobian to be non-singular.  Let us notice, more strongly, that (\ref{28}), written as
$\textbf{x}'-\hat{\textbf{r}}t'=\textbf{a}(\textbf{x},\textbf{0})$ implies that $\textbf{a}(\textbf{x},\textbf{0})$ is nonsingular
for three of the primed variables to be independent, and then also $g_{t}(\textbf{x},t)\neq 0$, for all four of the primed variables
to be independent.\\

Let us then define

\begin{equation}
\gamma \equiv g_{t}(\textbf{0},0)\neq 0. \label{gam}
\end{equation}

Subtracting (\ref{28}) from (\ref{27}), we have for the scalar function $g$,

\begin{equation}
(\textbf{r}(\textbf{v})- \hat{\textbf{r}})g(\textbf{x},t)=
\textbf{a}(\textbf{x},\textbf{0})- \textbf{a}(\textbf{x}-
\textbf{v}t, \textbf{v}). \label{29}
\end{equation}

(\ref{29}) is also correct for all \textbf{v}, \textbf{x}, $t$.  Substitute $\textbf{x}\rightarrow \textbf{x}+\textbf{v}t$, so that the second
\textbf{a} is independent of time, and differentiate on $t$.  Now, re-substitute $\textbf{x}\rightarrow \textbf{x}-\textbf{v}t$ to obtain

$$(\textbf{r}(\textbf{v})- \hat{\textbf{r}})(\textbf{v}\cdot \partial +\partial_{t})g(\textbf{x},t)= (\textbf{v}\cdot
\partial)\textbf{a}(\textbf{x},\textbf{0}),$$
or,

\begin{equation}
\textbf{r}(\textbf{v})- \hat{\textbf{r}}=\frac{(\textbf{v}\cdot
\partial)\textbf{a}(\textbf{x},\textbf{0})}{(\textbf{v}\cdot \partial +\partial_{t})g(\textbf{x},t)}. \label{29t}
\end{equation}
But, the RHS must be totally independent of both \textbf{x} and $t$, so that we may equate the RHS to its evaluation at $\textbf{x}=\textbf{0}$
and $t=0$.  With $\gamma \neq 0$, we can write this, in component form, as
\begin{equation}
\frac{(\textbf{v}\cdot
\partial)a_{i}(\textbf{x},\textbf{0})}{(\textbf{v}\cdot \partial +\partial_{t})g(\textbf{x},t)}=\frac {M_{ij}v_{j}}
{1-\textbf{b}\cdot\textbf{v}}, \label{29tt}
\end{equation}
with the linear operator \textbf{M} and the vector \textbf{b} constants, and by the comments following (\ref{28}), \textbf{M} is
invertible. This relation holds for all \textbf{v}, and so (\ref{29tt}) first must agree in its terms linear in \textbf{v}, yielding

\begin{equation}
\partial_{j}a_{i}(\textbf{x},\textbf{0})=g_{t}(\textbf{x},t)M_{ij}. \label{v1}
\end{equation}
With \textbf{M} invertible, at least one of the $M_{ij}\neq 0$.  But then, by (\ref{v1}),

\begin{equation}
g_{t}(\textbf{x},t) \textrm{ is independent of }t. \label{v2}
\end{equation}

But now, (\ref{29tt}) has the further content that

\begin{equation}
\partial g(\textbf{x},t)=-g_{t}(\textbf{x},t)\textbf{b}, \label{v3}
\end{equation}
for the denominators to agree.  The general solution to (\ref{v3}) is $$g(\textbf{x},t)=u(t-\textbf{b}\cdot\textbf{x}),$$ for an arbitrary
function $u$.  By (\ref{v2}) this function must be linear, which then by (\ref{gam}), has fully determined $g$:
\begin{equation}
t'=g(\textbf{x},t)=\gamma (t-\textbf{b}\cdot\textbf{x}),\;\gamma,\:\textbf{b}\textrm{ constants}. \label{34}
\end{equation}

But then, (\ref{v1}) reads $$\partial_{j}a_{i}(\textbf{x},\textbf{0})=\gamma M_{ij},$$ and so we have determined that
\begin{equation}
\textbf{a}(\textbf{x},\textbf{0})=\gamma \textbf{M}\cdot\textbf{x}. \label{v5}
\end{equation}

Finally, substituting (\ref{34}) and (\ref{v5}) in (\ref{28}), and renaming $\hat{\textbf{r}}\rightarrow -\textbf{A}$, \textbf{f} is linear in
\textbf{x} and $t$:

\begin{equation}
\textbf{x}'= \textbf{f}(\textbf{x},t)=  \gamma(\textbf{L} \cdot
\textbf{x} - \textbf{A}t), \; \textrm{vector \textbf{A} and linear
transformation \textbf{L} constant.} \label{35}
\end{equation}\

It should be noted that an arbitrary integration constant can be added to both (\ref{34}) and (\ref{35}), so that, more precisely, we
have shown the transformations to be affine.  These constants are precisely the freedom to arbitrarily choose an origin in space and
time.  It is this freedom that is usually understood as ``homogeneity", and invariance under it utilized in the variable $\xi$ and
the function $\textbf{a}(\xi,\textbf{v})$ of the hypothesis.  This is not sufficient to deduce that the relations are affine. Rather,
that $\textbf{r}(\textbf{v},\xi)$ is independent of $\xi$ is the critical extra ingredient to reduce what would have been projective
to the circumstance of affine.  We have no need for the sequel to consider these constants, and so retain just the homogeneous, or linear form.\\

We have thus proven that Galilean spaces are related, by (\ref{34}) and (\ref{35}), by linear transformations in both space and time. Before
fully determining their form, in consequence of relativity, part II, we need to comment on $t'$, when \textbf{b}$\neq$\textbf{0} in (\ref{34}).

%%%%%%%%%%%%%%%%%%%%%%%%%%%%%%%%%%%%%%%%%%%%%%%%%%%%%%%%

\section{Galileo's Time}

For a world possessing large numbers of natural uniform motions, one defines time as their consensus. To perform his experiments on non-uniform
motion down inclined planes (the latter a device to proportionally slow things down), Galileo needed to reliably measure shorter intervals of
time then had ever previously been attempted. His clock was a very broad tank of water with a small orifice that he manually opened and closed.
The efflux was gathered in a calibrated beaker. The presumption - to be cross-checked with other uniform processes - was that with a very small
drop in the level of fluid in the tank, the efflux would be uniform. This served as a ``master" time-piece with all others measured by it. To
measure a uniform velocity, one measures the time of passage between ruled lengths. The reference clock is good if, for a given uniform speed of
motion, the clock measures times exactly proportional to the lengths traversed. And this must be just as true for any - and all - other speeds of
uniform motion. It is only this cross-checked consensus that grants a meaning to time. It, in turn, is predicated on the large supply of
\textit{natural} motions, and it is the simple temporal description of these that confers a metrization upon time. Without the sure belief in
inertial uniform motion, there would be no candidates for consensus and calibration. It was precisely with this request to have a ``time" in the
back of mind that I started in section I with the two arguments Galileo adduced for inertial
motion. \\

The inclined plane was a table-top experiment, and so one ``clock" with his finger on it sufficed within his near field of vision to provide
time. The verification of the uniform motion of a large ship would have requested remote clocks as well. These separate clocks needed to have
been calibrated against each other, and then synchronized to be able to measure a velocity by subtracting the recorded time of one clock from
that of a remote clock. We proceed by having the clocks next to each other, having meticulously recorded how one constructed them, and then just
how we adjusted them to have them agree in  their rates. A ``remote" clock is now transported to its intended remote location. With sloshing of
water in the tank, or what not, it is at least likely that the rates will disagree during transport, but then agree when placed at rest in its
final location. So, we must find a means of synchronizing them when mutually remote. \\

Before doing so, I give the world V a copy of my meticulously recorded notes, and instruct him to assemble his remote clock-work, following the
notes to the letter. Should for some reason V discover that the notes didn't work for him, then his world is \textbf{not} identical to mine, and
Galileo's conception refuted. So, in the worlds of Galileo, V has an
identical set of clocks to mine. \\

Now I synchronize. I do so by the meaning of time as a consensus of any and all natural motions. Consider two identical massive balls compressing
a light spring. By \textbf{isotropy} these, when released, must fly off with exactly equal, but opposite speeds, with which they then continue,
inertially, to move at. It is of no importance to know what this speed is. With differing equal masses and differing springs, I can do this with
a vast diversity of speeds, with their equal and opposite directions, by isotropy, oriented at will. These are the calibrating natural
motions in my worlds. \\

Let us proceed. Between the first pair of remote clocks I choose to synchronize, I pace out at my leisure and stand exactly equidistant between
the two. Both clocks are off, but set to read midnight. I release an equal and opposite pair from the midpoint. Since equal distances are
traversed in equal times - by the definition of time - each clock is hit by its respective ball at the same time, and the reception of the ball
is the signal to start the clock. The two are now synchronized - without any knowledge of the exact speeds $\pm v$ of the pair of balls. To see
that time is indeed time, I proceed to repeatedly send off pairs with all different values of $\pm v$, and have both clocks keep a record of the
observed hitting times. ``Time" is time if
the two clocks have records of always identical hitting times. \\

To synchronize a third clock, stand midway between it and the nearer of the synchronized pair, and again release a $\pm v$ pair. The new clock is
set to read midnight and is off, starting up upon receiving a ball. The already working clock just records its time at the moment of impact of
the other ball. Now, at leisure, we bring the recording to the third clock, and advance its reading by exactly the recorded time. And so,
evidently, all clocks can be mutually
synchronized. \\

It is not accidental that we walk over with recorded values. On reflection, it is clear that time can only have meaning when space, indelibly and
fully objectively, can be altered to record past instants. Time lives inherently deeply tied to
properties of space. \\

The world of V synchronizes itself in this identical way, and cross-checks by timing different $\pm v$'s for a diversity
of $v$'s. \\

There is \textit{nothing} about this that requires light, with its peculiar properties, and a special speed $c$. Should it exist, we must, of
course, cross-check our clocks that they are simultaneous for a $\pm c$ pair. But the existence of light is neither necessary \textit{nor}
sufficient to guarantee that time is ``time". ``Time" exists only by consensus among natural motions. Light has no special role, whatever, in the
synchronizing of time, save for a refined technology, should that technology be
extant. \\

Now we reach the crux of the matter and realize that for V and I to both have time, then a point transformation, II.\ref{23} is
\textit{mandatory}. To streamline the requisite experimental test, we capitalize on the homogeneity of time and space to place one of my
synchronized clocks at $\textbf{x}=t=0$ and the other say at \textbf{x}=$2L\hat{\textrm{\textbf{x}}}$, and sit at
\textbf{x}=$L\hat{\textrm{\textbf{x}}}$. With more skillful preparation I can produce $\pm v$ pairs at a diversity of \textit{known} $v$'s. I
send out my slowest pair first at time $t= -L/v_{<}$. I repeatedly launch each pair, slower before faster, at the respective times $-L/v$,
launching the last, the fastest, at $t= -L/v_{>}$. By contrivance, \textit{all} the $-v$ balls hit \textbf{x}=\textbf{0} precisely at $t=0$, and
\textit{all} the $+v$ balls hit \textbf{x}=$2L\hat{\textrm{\textbf{x}}}$ again at $t=0$. \\

What does V see? Should he see the balls hitting my \textbf{x}=\textbf{0} clock at all different times, he would conclude that something is
profoundly wrong with my clock because it fails to register equal distances proportional to speed. Either I really have nothing that is ``time"
or he doesn't. In either case our worlds are not equivalent and the entire conception of natural motions deeply erroneous. So, in the equivalent
worlds of Galileo with clocks of his conception, it must be that V also sees all the balls hitting my clock at \textbf{x}=\textbf{0} at some one
instant of his time $t'$. But this says $t'=g(\textbf{x},t)$: to a unique point \textbf{x}=\textbf{0} \textit{and} time, $t=0$, corresponds
precisely one $t'$ as seen by V. And just so, $\textbf{x}'=\textbf{f}(\textbf{x},t)$: to the unique $(\textbf{x},t)$ corresponds a unique
\textbf{x}$'$. To have consistent time between two worlds there can be no choice but for a point transformation, II.\ref{23}, which moreover, as
we learned in
II.\ref{34}, \ref{35}, is linear. \\

However, while V must then also see all the balls hitting my clock at \textbf{x}=$2L\hat{\textrm{\textbf{x}}}$, $t=0$, if \textbf{b} of
II.\ref{34} is not orthogonal to $\hat{\textrm{\textbf{x}}}$, then rather than seeing all of them hit at $t'=0$ for the clock at
$(\textbf{x}=\textbf{0}, t=0)$, instead at some different time $t'\neq 0$. While V might have opined for a simpler, tidier set of
world-relations, there is nothing, however, inconsistent about such an outcome: One can only meticulously
observe the world to discover what it is. \\

In all cases, this is time as Galileo might have conceived it, in no ways requiring the special and unusual properties of
light, of which, of course, he was ignorant.  \\

%%%%%%%%%%%%%%%%%%%%%%%%%%%%%%%%%%%%%%%%%%%%%%%%%%%%%%%%

\section{The World Systems Of Galileo: Lorentz Transformations With Some Constant of Nature $1/c^{2}$}

Let us now marry the relativity of II to the homogeneous, isotropic and linearly related worlds of III. This will produce all the systems that
are compatible with all of Galileo's thoughts about motion. The role of isotropy is critical, and here rises to the fore. It appears in all
treatments, after the prior annunciation of ``the postulates" as a calling upon ``symmetry". Let us be totally explicit, since in the usual
treatments, references to isotropy are not elucidated. If I
need assume it, I rather assume it all. \\

With $\textbf{x}= \textbf{v}t$ in (\ref{34}) and (\ref{35}), we produce the rule \textbf{r} of (\ref{1}):

\begin{equation}
\textbf{r}(\textbf{v},\textbf{V})= \textbf{v}'= \frac{\textbf{L}
\cdot \textbf{v}-\textbf{A}}{1-\textbf{b} \cdot \textbf{v}} .
\label{36}
\end{equation}

To identify parameters, by II.\ref{17}, \textbf{r}(\textbf{0},\textbf{V})= -\textbf{V}, and so with \textbf{v}=\textbf{0} in (\ref{36}),

\begin{equation}
\textbf{A}=\textbf{V}, \label{37}
\end{equation}

and one vector of parameters is determined. (Recall here that with II.\ref{17}, we are seeking transformations to identically oriented worlds
only.) Next, by II.\ref{2}, \textbf{r}(\textbf{V},\textbf{V})=\textbf{0}, so by (\ref{36}),

\begin{equation}
\textbf{L} \cdot \textbf{V}=\textbf{V}, \label{38}
\end{equation}

and the linear transformation \textbf{L} has \textbf{V} as a right
eigenvector of eigenvalue +1. \\

Next we appeal to isotropy. My world is totally isotropic. The rule \textbf{r}, by (\ref{36})-(\ref{38}) has explicitly encoded a particular
direction $\widehat{\textbf{V}}$, which breaks the full isotropy. However, there is no physical entity or cause appearing that will further break
the residual, polar isotropy about the axis $\widehat{\textbf{V}}$: Our relations are required to preserve
full polar isotropy about $\widehat{\textbf{V}}$. \\

Consider now (\ref{34}) and (\ref{35}). Since $\widehat{\textbf{V}}$ can be chosen arbitrarily, (\ref{35}) is a rotational vector equation and
(\ref{34}) a rotational scalar. \textbf{b} must then be a rotational vector, and $\gamma$ a rotational scalar. That is, upon an arbitrary
rotation of the two spatial worlds by \textbf{R}, by isotropy, since $t'$ must be invariant,
$$t'_{R}= \gamma(\textbf{R} \cdot \textbf{V})
(t-\textbf{b}(\textbf{R}\cdot\textbf{V}) \cdot \textbf{R} \cdot\textbf{x}), $$
$$t'= \gamma(\textbf{V})(t- \textbf{b}(\textbf{V}) \cdot
\textbf{x}) = \gamma(\textbf{V})(t-(\textbf{R} \cdot \textbf{b}(\textbf{V})) \cdot (\textbf{R} \cdot\textbf{x})),$$ and so, since $t'_{R}=t'$,

\begin{equation}
\gamma(\textbf{R} \cdot \textbf{V}) = \gamma (\textbf{V}), \textrm{
and} \label{39}
\end{equation}

\begin{equation}
\textbf{b}(\textbf{R} \cdot \textbf{V})= \textbf{R} \cdot
\textbf{b}(\textbf{V}). \label{40}
\end{equation}

By a suitable rotation, \textbf{R}, whatever \textbf{V} is, I can rotate it to $|\textbf{V}|\hat{\textbf{x}}$, and so by (\ref{39}),

\begin{equation}
\gamma(V_{1}, V_{2}, V_{3}) = \gamma(|\textbf{V}|,0,0)\equiv
\gamma(V^{2}) \label{41}
\end{equation}

That is, of course, to be a rotational scalar, $\gamma$ can depend upon \textbf{V} only through \textbf{V}'s modulus, or $V^{2}$. As for
\textbf{b}, it is either the constant \textbf{0}, or a rotational vector. The only one at hand is \textbf{V} itself. The declaration of any
other, orthogonal to \textbf{V}, will break polar isotropy, which we may never to do unless an explicit physical cause
forces us to do so. Thus, \\

\begin{equation}
\textbf{b}(\textbf{V})= \frac{\textbf{V}}{c^{2}(V^{2})} \label{42}
\end{equation}

with $1/c^{2}(V^{2})$ again a rotational scalar.
\\

At this point we have $t'$ of (\ref{34}),

\begin{equation}
t'= \gamma(V^{2})(t- \frac{\textbf{V} \cdot
\textbf{x}}{c^{2}(V^{2})}). \label{43}
\end{equation}

What is most important is that $\textbf{V} \cdot \textbf{x}=\textbf{V} \cdot
 \widehat{\textbf{V}}  ( \widehat{\textbf{V}}   \cdot  \textbf{x})+\textbf{V}  \cdot  \textbf{x} _{\perp}$, or $\textbf{V}
\cdot \textbf{x}=\textbf{V} \cdot \widehat{\textbf{V}}
(\widehat{\textbf{V}} \cdot \textbf{x})$ since $\textbf{V}\perp
\textbf{x}_{\perp}$. Had \textbf{b} been other than along
$\widehat{\textbf{V}}$, then V's time would vary with polar
rotations about $\widehat{\textbf{V}}$, and V's ``up"
time would disagree with his ``down" time. \\

As for (\ref{35}), recalling (\ref{37}) and(\ref{38}),

\begin{equation}
\textbf{x}'=\gamma(V^{2})(\textbf{L} \cdot \textbf{x}-\textbf{V}t);
\; \textbf{L} \cdot \textbf{V}=\textbf{V}, \label{44}
\end{equation}

and (\ref{36}) reads,

\begin{equation}
\textbf{r}(\textbf{v}, \textbf{V})= \frac{\textbf{L} \cdot
\textbf{v} - \textbf{V}}{1- \textbf{V} \cdot
\textbf{v}/c^{2}(V^{2})} . \label{45}
\end{equation}

Before determining \textbf{L}(\textbf{V}), we already have an extraordinary result in consequence of II.(\ref{20}) of
relativity. \\

Consider all those \textbf{v}'s parallel to $\widehat{\textbf{V}}$, writing

\begin{equation}
\textbf{v}\equiv v \widehat{\textbf{V}}, \; \textbf{V}=V
\widehat{\textbf{V}}. \label{46}
\end{equation}

Since $\textbf{L}\cdot \widehat{\textbf{V}}=\widehat{\textbf{V}}$, we have

$$\textbf{r}(v \widehat{\textbf{V}},V \widehat{\textbf{V}})=
\widehat{\textbf{V}} \frac{v-V}{1- Vv/c^{2}(V^{2})}\equiv \widehat{\textbf{V}}r(v,V) ,$$

or

\begin{equation}
r(v,V) = \frac{v-V}{1- Vv/c^{2}(V^{2})}. \label{47}
\end{equation}

But, with \textbf{v} and \textbf{V} parallel, we have by (\ref{20}),

\begin{equation}
r(v,V)= -r(V,v) = \frac{v-V}{1- Vv/c^{2}(v^{2})}. \label{48}
\end{equation}

Comparing (\ref{48}) to (\ref{47}) we now conclude

\begin{equation}
c^{2}(v^{2})=c^{2}(V^{2}), \; \textrm{all} \; v^{2}. \label{49}
\end{equation}

By rotational isotropy, the parameter $c^{2}$ depends only upon the modulus of \textbf{V} of the world V, and is some unknown function of its
argument. By (\ref{49}), for every value of $v^{2}$, $c^{2}$ is unchanged, and so $c^{2}$ is a \textit{constant} function. That is

\begin{equation}
1/c^{2}(V^{2}) \equiv  1/c^{2} \label{50}
\end{equation}

for some one unique value of $1/c^{2}$ common to all worlds. That is, there is some universal constant of nature, and each possible system of
worlds obeying all of Galileo's concepts, is a system of a fixed value of $1/c^{2}$. $1/c^{2}$ in particular could be precisely 0. Galileo's
thoughts can't determine this;
only experiment can. \\

Notice, there was never a need to invoke light to have the scalar $c^{2}(V^{2})$ reduce to a pure constant. This was already implicit in the
assumptions of relativity, homogeneity and isotropy. This is the end of the ``$2^{nd}$ Postulate". It risked only
being false - \textit{i.e.} incompatible with the first. \\

Let us rewrite (\ref {43})-(\ref{45}) using (\ref{50}), and then go on to determine \textbf{L} and $\gamma$.

\begin{eqnarray}
\textbf{x}' &=& \gamma(V^{2})(\textbf{L} \cdot \textbf{x} - \textbf{V}t), \label{51} \\
t' &=& \gamma(V^{2})(t - \textbf{V} \cdot \textbf{x}/c^{2}), \; \textrm{and} \label{52} \\
\textbf{v}' &\equiv &\textbf{r}(\textbf{v},\textbf{V}) = \frac{\textbf{L} \cdot \textbf{v}-\textbf{V}}{1-\textbf{V} \cdot \textbf{v}/ c^{2}}
\label{53}
\end{eqnarray}

By isotropy, \textbf{L}(\textbf{V}) is a rotational 2-tensor:

$$\textbf{R} \cdot \textbf{x}'= \gamma(\textbf{L}(\textbf{R} \cdot \textbf{V}) \cdot \textbf{R}\textbf{x}-(\textbf{R}\textbf{V})t)$$

$$\gamma \textbf{R} \cdot (\textbf{L}(\textbf{V}) \cdot \textbf{x}
- \textbf{V}t) = \gamma (\textbf{R} \cdot \textbf{L}(\textbf{V}) \cdot \textbf{R}^{-1} \cdot \textbf{R}\textbf{x} -(\textbf{R}\textbf{V})t),$$
and so
\begin{equation}
\textbf{L}(\textbf{R} \cdot \textbf{V}) = \textbf{R} \cdot \textbf{L}(\textbf{V}) \cdot \textbf{R}^{-1} \label{54}
\end{equation}

Here again, to introduce no new vector orthogonal to \textbf{V}, thereby maintaining polar isotropy,

\begin{equation}
\textbf{L}(\textbf{V}) = \alpha(V^{2})\widehat{\textbf{V}} \widehat{\textbf{V}} + \beta(V^{2})(\textbf{1} - \widehat{\textbf{V}}
\widehat{\textbf{V}}). \label{55}
\end{equation}

(Let us point out, that under isotropy, we can also add to \textbf{L} of (\ref{55}) $\delta (V^{2}) \frac{\textbf{V}}{c}\wedge$. This
antisymmetric part, when subjected to the following treatment, produces simply a rotation about $\widehat{\textbf{V}}$ performed
\textit{after} the complete determination of the form of (\ref{51}). Since, at this point, we are requesting a transformation with
\textit{parallel} axes, we require $\delta=0$, and then finally reinstate it, amongst all possible rotations of V's axes at the end.
%Indeed, under parallelism, we noted above (\ref{acom}) that $-\textbf{r}(\textbf{v},\textbf{V})=\textbf{r}(-\textbf{v},-\textbf{V})$.
%Together with (\ref{53}), we then conclude that $\textbf{L}(- \textbf{V})= \textbf{L}(\textbf{V})$.  But then, under parallelism, we
%demand $\delta (V^{2})=0$.
Indeed, by parallelism,  explained above (\ref{2}), \textbf{r},\textbf{v}, and \textbf{V}
are coplanar, so that $\delta (V^{2})$ must vanish.)\\

[For the reader less at ease with the idea that only \textbf{V} is at hand, the form of \textbf{L}(\textbf{V}) of (\ref{55}) is
deduced as follows.  Consider (\ref{54}) first for the abelian subgroup of rotations about the \textbf{V}-axis, which leave
\textbf{V} invariant.  (\ref{54}) now says all such 2-D rotations commute with \textbf{L}(\textbf{V}).  Write down a general 3x3
matrix for \textbf{L} that leaves the x-axis invariant, and write down the consequence of its commutation with the 3x3 that is just a
$\theta$ rotation in the y-z plane.  This results in (\ref{55}) with the $\delta$ term as well.  Each of the coefficients is a
function of \textbf{V}.  Now apply (\ref{54}) for \textit{any} \textbf{R}, and discover that each is a rotational scalar.]\\

That is, \textbf{L} must polar decompose, with rotationally invariant coefficients in the 1-D polar space, $\widehat{\textbf{V}}
\widehat{\textbf{V}}$, and in the 2-D $\perp$ space $\textbf{1}- \widehat{\textbf{V}} \widehat{\textbf{V}}$. In addition, since $\textbf{L} \cdot
\widehat{\textbf{V}} = \widehat{\textbf{V}}$, $\alpha \equiv1$:

\begin{equation}
\textbf{L}(\textbf{V}) = \widehat{\textbf{V}} \widehat{\textbf{V}} + \beta(V^{2})(\textbf{1}- \widehat{\textbf{V}} \widehat{\textbf{V}}).
\label{56}
\end{equation}

In particular,

\begin{equation}
\textbf{L}(- \textbf{V})= \textbf{L}(\textbf{V}), \label{57}
\end{equation}

and we are ready to use relativity, II.\ref{22} to invert (\ref{53}) for \textbf{v}:

\begin{equation}
\textbf{v}= \frac{\textbf{L} \cdot \textbf{v}' + \textbf{V}}{1+ \textbf{V} \cdot \textbf{v}'/ c^{2}  }. \label{58}
\end{equation}

Substituting (\ref{53}) into (\ref{58}), obtain the identity for \textbf{v}, using $\textbf{L} \cdot \textbf{V}=\textbf{V}$,

\begin{equation}
\textbf{v} = \frac{(\textbf{L}^{2} - \textbf{V} \textbf{V} / c^{2}) \cdot \textbf{v}}{(1- V^{2}/c^{2})+ (\textbf{V} \cdot \textbf{L} -
\textbf{V}) \cdot \textbf{v}/c^{2}}. \label{59}
\end{equation}

With \textbf{v} arbitrary, we have

\begin{equation}
\textbf{V} \cdot \textbf{L} = \textbf{V}, \; \textrm{and} \label{60}
\end{equation}

\begin{equation}
\textbf{L}^{2} = \frac{\textbf{V}\textbf{V}}{c^{2}}+(1- V^{2}/c^{2})\textbf{1} = \widehat{\textbf{V}} \widehat{\textbf{V}} +
a^{2}(V^{2})(\textbf{1}- \widehat{\textbf{V}} \widehat{\textbf{V}}), \; \textrm{with} \label{61}
\end{equation}

\begin{equation}
a^{2}(V^{2})=1- V^{2}/c^{2}. \label{62}
\end{equation}

Now $\textbf{L} \cdot \textbf{V} = \textbf{V}$ with (\ref{60}) $\textbf{V} \cdot \textbf{L} = \textbf{V}$ imply that \textbf{L} splits into a
direct sum in $\widehat{\textbf{V}} \widehat{\textbf{V}}$ and its perp-space, $\textbf{1}- \widehat{\textbf{V}} \widehat{\textbf{V}}$ which we
already determined in (\ref{54}). Comparing (\ref{61}) and (\ref{56}), we know \textbf{L} completely:

\begin{equation}\textbf{L}(\textbf{V}) = \widehat{\textbf{V}} \widehat{\textbf{V}} + a(V^{2})(\textbf{1}- \widehat{\textbf{V}} \widehat{\textbf{V}}) \;
\textrm{with} \label{63}
\end{equation}

\begin{equation}
a(V^{2})= + \sqrt{1-V^{2}/c^{2}}, \label{64}
\end{equation}

where the + root is taken so that as $V \rightarrow0$, $\textbf{L} \rightarrow \textbf{1}$  to agree with II.(\ref{3}). In particular the rule
\textbf{r} of Galilean worlds II.(\ref{1}) is fully determined by (\ref{53}), and we now know \textit{all} the world systems compatible with
Galileo, he having had the conceptual and mathematical prowess to determine the one system $1/c^{2}=0$.
\\

All that remains to fully determine the transformation rules is $\gamma(V^{2})$. Project (\ref{51}) on any direction $\perp$ to
$\widehat{\textbf{V}}$, and obtain by (\ref{63})

\begin{equation}
x'_{\perp} = \gamma(V^{2})a(V^{2})x_{\perp}. \label{65}
\end{equation}

We now employ relativity for the last time, with II.(\ref{17}). (Indeed, this is the only need we have for the full statement of
inertial relativity, since up to here we have only utilized it with respect to velocities.)  V sees me moving at -\textbf{V},
(\ref{65}) depends only upon $V$, and so V must conclude

\begin{equation}
x_{\perp} = \gamma(V^{2})a(V^{2}){x_{\perp}}'. \label{66}
\end{equation}

But then, not to induce a $\pi$ rotation about $V$ as $V \rightarrow 0$,

\begin{equation}
\gamma(V^{2})= \frac{1}{a(V^{2})} \;, \label{67}
\end{equation}

and (\ref{51}) and (\ref{52}) are fully determined.
\\

Finally, (\ref{51}) can be generalized by allowing V to now arbitrarily reorient his axes by some rotation \textbf{R}:

\begin{equation}
\textbf{x}'= \textbf{R} \cdot \gamma(V^{2})(\textbf{L} \cdot \textbf{x} - \textbf{V}t) \; , \; \textrm{any rotation}, \;
\textbf{R}^{t}\textbf{R}=\textbf{1}. \label{68}
\end{equation}

(\ref{68}), (\ref{52}), (\ref{63}), (\ref{64}), and (\ref{67}) constitute precisely the Lorentz group at parameter $1/c^{2}$. These, and just
these, are the possible world systems implied by Galileo's thoughts made mathematical. There is
no conceptual role \textit{anywhere} for light to enter. \\

While \textbf{R} in (\ref{68}) might appear as a why-not luxury, no such thing is true for $1/c^{2}\neq0$. It is straightforward, but somewhat
tedious, to verify the cautionary provisos that $\textbf{R}(\textbf{V}',\textbf{V})$ of (\ref{8}) and (\ref{18}) is not the identity when
$1/c^{2}>0$. In particular, parallelism is not transitive and $\textbf{R}(\textbf{V}',\textbf{V})$ is a rotation of angle $\theta$ about axis
$\hat{\textbf{n}}$ with

\begin{equation}
\hat{\textbf{n}} \tan(\theta/2)=\frac{\textbf{V}'\wedge \textbf{V}/c^{2}}{(1+a)(1+a')-\textbf{V}'\cdot\textbf{V}/c^{2}}\: , \label{69}
\end{equation}

with $$ a=a(V^{2}), \: a'=a(V'^{2})\; \textrm{of }(\ref{64}),$$

which, to $O(1/c^{2})$ is,

$$\textbf{R}(\textbf{V}',\textbf{V})=\textbf{1}+ \frac{\textbf{V}'\wedge \textbf{V}}{2c^{2}}\wedge \;+\; \ldots\ $$
\\
At this point let us fill in a few elementary points.  When we say $1/c^{2}\neq0$, we actually mean $1/c^{2}>0.$  For example, with parallel
velocities, (\ref{47}), after the result of  (\ref{50}) is, with $1/c^{2}<0$, simply the tangent of the difference of two angles, where $v/c$ is
the tangent of an angle.  Thus, each successive application of the same transformation, subtracts the same angle, until, with an artfully chosen
transformation, we notice that after $n$ applications of it, we have the finite $v$ infinite.  This is quite alien to notions of energy
conservation.  Pursuing the thought, we realize that with $1/c^{2}<0$, the space-time transformation itself is just a rotation in the plane of
the spatial direction of $\widehat{\textbf{V}}$ and $t$.  So, just as $v$ becomes infinite, this direction of space exchanges roles with time.
This is certainly not an equivalent Galilean world to the one we started in, and so we may only accept $1/c^{2}>0.$\\
\\
Next, let us notice the most striking manner in which the two arguments of $\textbf{r}(\textbf{v},\textbf{V})$ have radically different
connotations.  The form of (\ref{53}) together with the projections of \textbf{L} of (\ref{63}),(\ref{64}) is a bit inconvenient to directly
employ.  With $\widehat{\textbf{V}}(\widehat{\textbf{V}}\cdot\textbf{v})\equiv \textbf{v}_{\|}$, notice that the numerator of (\ref{53}) contains
$\textbf{v}_{\|}$ in the form $(1-a)\textbf{v}_{\|}$.  But, by (\ref{64}), $(1-a)=V^{2}/(c^{2}(1+a))$, so that (\ref{53}) can be written as
\begin{equation}
\textbf{r}(\textbf{v},\textbf{V})=\frac{a}{1-\frac{\textbf{v}\cdot\textbf{V}}{c^{2}}}(\textbf{v}-\frac{\textbf{V}}{1+a})-\frac{\textbf{V}}{1+a},\label{r}
\end{equation}
with $a=a(V^{2})$.  Evidently, \textbf{v} and \textbf{V} enter (\ref{r}) in radically different ways.  Notice that with the first argument of
\textbf{r} at speed $c$, that is with $\textbf{v}=c\widehat{\textbf{v}}$, (\ref{r}) is perfectly well-behaved, with inverse, as usual, given by
(\ref{22}).  No such thing is true if we take $\textbf{V}=c\widehat{\textbf{V}}$ for the second argument of \textbf{r}, that is, for the velocity
of a \textit{world}.  Here, \textbf{r} is still well-behaved with $a=0$, and so
$$\textbf{r}(\textbf{v},c\widehat{\textbf{V}})=-c\widehat{\textbf{V}}.$$
Not only is this degenerate result not invertible, but, moreover, informs us that the object at $c$ is \textit{not} an equivalent Galilean world.
How limits are taken here matters, and the result, allowing for limits to $c$ from directions other than $\widehat{\textbf{V}}$ shows that there
is a profound asymmetry between \textbf{v} and \textbf{V}.  This means that not only is the 1-D story misleading, but actually wrong.  What is
the nature of this ``world" at $c$?  We can agree, in a choice of limits, that it contains a host of particles all at rest with respect to it.
But, every other particle's motion is at speed $c$ to it.  This world, then, has no means to build or see processes of uniform or periodic
motions:  It is a world with no time.  Or, it is a ``dead" world, with only stagnant neighbors in a sea of uncontrollably fast entities.  This is
certainly not an equivalent Galilean world, if indeed, one could construe its existence.\\
\\
That is, particles are allowed at $c$, but not worlds, and the order and meaning of the arguments of \textbf{r} is of critical importance.\\

Let me now conclude this section of marrying relativity with linearity with a rapid discussion of how $\textbf{R}(\textbf{V}',\textbf{V})$ is
related to the ``Wigner rotation" arising from the Lorentz group.  Our transformations, embodied in (\ref{70})-(\ref{71}), are those that
transform to a world where the particles I see with velocity \textbf{V} have been brought to rest, with axes aligned in both worlds.  The usual
object, a ``boost" $\textbf{B}_{\textbf{V}}$, is precisely the inverse, or our transformation for -\textbf{V}.  We pay no further attention to
this, but provide it as a dictionary entry for the reader who wants to convert between conventions.  However, what is usually meant by the Wigner
rotation is that the product of two non-parallel boosts is a Wigner rotation pre-multiplying the boost to the ``sum" of the two velocities.  We
leave it to the conversionary reader precisely where this rotation is to be applied, and just which choice of signs he cares about.  Whatever
that choice may be, the result, purely up to signs, is always \textit{precisely} (\ref{69}) in consequence of \textbf{R}'s structural
relationship (\ref{wig}).\\

Our \textbf{R} appears in our discussion in the context of the parametric structure \textbf{r}.  Let us now relate this to the group-structural
entity.  Our transformations have been rendered only in homogeneous (Lorentz) form, rather than the fuller inhomogeneous (Poincar\'{e}) form.  In
this homogeneous, linear form, the velocity, \textbf{v}, is precisely $\textbf{x}/t$, and the transformed velocity, $\textbf{v}'$,
$\textbf{x}'/t'$.  That is, \textbf{v} is transformed by the \textit{group} transformations as an induced projective transformation.  Let us now
call the space-time linear transformation (\ref{70})-(\ref{71}) $\textbf{T}_{\textbf{V}}$, so that we write
\begin{equation}
(\textbf{x}',t')=\textbf{T}_{\textbf{V}}\cdot (\textbf{x},t). \label{g1}
\end{equation}

Dividing by $t$ and $t'$, we rewrite (\ref{g1}) in its projective form
\begin{equation}
\textbf{T}_{\textbf{V}}\cdot (\textbf{v},1)=\mu (\textbf{v}',1), \label{g2}
\end{equation}

where the scaling $\mu$ stands for $t'/t$.  As used projectively, $\mu$ is determined by the lowest (time) component of (\ref{g2}), to then
determine the projective form of the upper (space) components.  This, of course, is precisely (\ref{72}).  But since $\mu$ is just a scalar
multiple, we can now determine the \textit{compostitions} of \textbf{v}'s by the successive group multiplications of the
$\textbf{T}_{\textbf{V}}$'s.  Accordingly, the master statement of relativity, (\ref{8}), is just the induced projective result of
\begin{equation}
\textbf{T}_{\textbf{r}(\textbf{V},\textbf{V}')}\cdot \textbf{T}_{\textbf{V}'}=\textbf{R}(\textbf{V}',\textbf{V})\cdot \textbf{T}_{\textbf{V}}.
\label{g3}
\end{equation}

While (\ref{g3}) suffices to determine the relation of \textbf{R} to any chosen definition of a ``Wigner rotation", we'll go slightly further, to
make it clear that whatever the definition, it is always exactly of form (\ref{69}).  Thus, set $\textbf{U}=\textbf{r}(\textbf{V}, \textbf{V}')$,
use (\ref{22}), replace $\textbf{V}'$ by \textbf{V}, and use \textbf{s} of (\ref{add}) in place of \textbf{r}  to obtain

$$\textbf{T}_{\textbf{U}}\cdot \textbf{T}_{\textbf{V}}=\textbf{R}(\textbf{V},\textbf{r}(\textbf{U},-\textbf{V}))\cdot
\textbf{T}_{\textbf{s}(\textbf{U},\textbf{V})}. $$ Finally, use (\ref{wig}) to obtain
\begin{equation}
\textbf{T}_{\textbf{U}}\cdot \textbf{T}_{\textbf{V}}=\textbf{R}(\textbf{U},-\textbf{V})\cdot \textbf{T}_{\textbf{s}(\textbf{U},\textbf{V})}.
\label{g4}
\end{equation}
That is, the fundamental \textbf{R} of (\ref{8}) is exactly a ``Wigner rotation", without any further computation, however the velocity arguments
are to be ``added" or ``subtracted".  These results are not obvious, but rather a part of the power of relativity.  If \textbf{r} is ``funny",
bearing no relation to a natural operation of either addition or multiplication, it is simply because it is projective.

\section {Kinematics And Dynamics: E=mc$^{2}$.}

We have, with (\ref{51}), (\ref{52}), (\ref{63}), (\ref{64}) and (\ref{67}) the transformations from one world to another with aligned axes, with
the corresponding formula (\ref{53}) for \textbf{r}(\textbf{v},\textbf{V}). To recapitulate in one place:

\begin{equation}
\textbf{x}'=\frac{\textbf{L} \cdot \textbf{x}- \textbf{V}t}{a(V^{2})} \; , \; t'= \frac{t-\textbf{V} \cdot \textbf{x}/c^{2}}{a(V^{2})} \label{70}
\end{equation}

with

\begin{equation}
\textbf{L}= \widehat{\textbf{V}} \widehat{\textbf{V}} + a(V^{2})(\textbf{1}- \widehat{\textbf{V}} \widehat{\textbf{V}}) \; , \; a(V^{2})=
{\sqrt{1-V^{2} /c^{2}}} \; . \label{71}
\end{equation}

(\ref{70}) produces \textbf{r}:

\begin{equation}
\textbf{r}(\textbf{v},\textbf{V})= \frac{\textbf{L} \cdot \textbf{v}- \textbf{V}}{1- \textbf{V} \cdot \textbf{v}/c^{2}}. \label{72}
\end{equation}

In full generality, with V reorienting his axes by a rotation \textbf{R},

\begin{equation}
\textbf{x}'\rightarrow \textbf{R} \cdot \textbf{x}' \; , \; t'\rightarrow t' \; , \; \textbf{r}(\textbf{v}, \textbf{V})\rightarrow \textbf{R}
\cdot \textbf{r}(\textbf{v}, \textbf{V}). \label{73}
\end{equation}

It is easy to check, by the simple polar decomposition of \textbf{L}, that

\begin{equation}
s^{2}\equiv(ct')^{2} - \textbf{x}'^{2}=(ct)^{2}- \textbf{x}^{2} \label{74}
\end{equation}

for all transformations (\ref{73}), and moreover, that (\ref{73}) exhausts the group of linear transformations
that has $s^{2}$ invariant.  \\

Is is equally straightforward to determine from (\ref{72}) that

\begin{equation}
\frac{1}{\sqrt{1- r^{2}(\textbf{v},\textbf{V})/c^{2}}}= \frac{1-\textbf{v} \cdot \textbf{V}/c^{2}}{a(v^{2})a(V^{2})}\; , \label{75}
\end{equation}

a result we're about to capitalize upon. Notice, however, that (\ref{75}) is symmetric in $\textrm{\textbf{v}} \leftrightarrow
\textrm{\textbf{V}}$, so that we have verified our foundational relativity relation II.\ref{19}, with \textbf{R} of II.\ref{18} just a rotation.
Indeed, given (\ref{75}), and with an easy similar calculation for $1- \widetilde{\textbf{v}} \cdot \widetilde{\textbf{V}}/c^{2}$, we see that
the Principle of Relativity, II.\ref{7} and II.\ref{8} are also correct for any one value of $1/c^{2}$: there can be no conceptual
way to decide the value of $1/c^{2}$. \\

We now want to determine the kinematics of particles implied by (\ref{70})-(\ref{73}). We know we can do so, based on the brilliant idea of
Einstein's 1905 deduction of $M=E/c^{2}$. The argument implemented by Einstein rests on the isotropic emission of a $\pm c$ pair of finite
volumes of light. As we gather from section IV, $\pm \textrm{\textbf{v}}$ masses, but
with all possible v's should also suffice. Let us proceed. (This section is a streamlined version of \cite{11}, which I reproduce here to show
that the \textit{entire} construction of special relativity is unreliant on light, and indeed, quite simply so.)      \\

We consider a large symmetric mass at rest. A narrow cylinder with axis through the center of the mass has been bored out. We prepare a pair of
small equal masses compressing a light spring and put the assembly centered in the cylindric bore. This composite entity, at rest, is our initial
object. By isotropy, we can orient the axis of the bore at will. At some point we release the pair, which then by isotropy, fly off at equal and
opposite velocities $\pm \textrm{\textbf{v}}$ \textit{and}, throughout the emission process, the large mass remains at rest. This remaining large
object is in a new state - for example, it's lost the mass of the emitted pair. However, with the pair gone, by isotropy,
this final state is independent of the emission direction. \\

We also watch this from a world V, moving say at velocity $-V \hat{\textbf{x}}$. To V, the initial state is the one large object moving uniformly
at $+V\hat{\textbf{x}}$. It persists to move at $+V \hat{\textbf{x}}$ during and after the emission process. The emitted entities appear at the
non-equal speed

\begin{equation}
\textbf{v}'_{\pm} = \textbf{r}(\pm \textbf{v}, -V \hat{\textbf{x}}) \; . \label{76}
\end{equation}

By polar isotropy about $\hat{\textbf{x}}$, it suffices to consider the plane containing $\hat{\textbf{x}}$ and $\pm \textrm{\textbf{v}}$. We
call this other axis of the plane $\hat{\textbf{y}}$, so that

\begin{equation}
\pm \textrm{\textbf{v}}= \pm |v|(\cos \alpha, \sin \alpha) \label{76pm}
\end{equation}

with $\hat{\textbf{v}} \cdot \hat{\textbf{y}} = \sin \alpha$.

We now ask if there is a scalar conserved additive quantity, depending on the \textit{speed} ($|\textbf{v}|$) alone of each entity (i.e.
rotational scalars, by isotropy). Let us call this

\begin{equation}
k(v^{2}) \label{77}
\end{equation}

for each of the small masses of the emitted pair, and \\

\begin{equation}
K_{i,f}(V^{2}) \label{78}
\end{equation}

for the same quantity for the initial and final states of the large object as viewed by V. The only difference of that object as viewed from V,
as opposed to me at $K_{i,f}(0)$ is just whatever extra energy the large objects acquires by virtue of its possessing speed, $|\textbf{V}|$. So,
we take, by unimportant convention,

\begin{equation}
K_{i,f}(0)=k(0)\equiv 0 \;, \label{79}
\end{equation}

so that (\ref{78}) is precisely the \textit{kinetic energy} (by
definition) of the large particle. \\

So, we write down the balance of the sums of these energies in V and in me and subtract the two:

\begin{equation}
K_{i}(V^{2})= K_{f}(V^{2})+ k({v'_{+}}^{2})
+k({v'_{-}}^{2})-2k(v^{2}) \; . \label{80}
\end{equation}

Given the simplicity of (\ref{75}), let us define

\begin{equation}
k(v^{2}) \equiv u(\frac{1}{\sqrt{1-v^{2}/c^{2}}}) \; . \label{81}
\end{equation}

Then, by (\ref{75}), (\ref{76}) and (\ref{76pm}), (\ref{80}) becomes

\begin{equation}
K_{i}(V^{2})- K_{f}(V^{2}) = u(\frac{1 + vV \cos \alpha/c^{2}}{\sqrt{1-V^{2}/c^{2}} \sqrt{1- v^{2}/c^{2}}})+u (\frac{1- vV \cos \alpha
/c^{2}}{\sqrt{1-V^{2}/c^{2}} \sqrt{1- v^{2}/c^{2}}})- 2u(\frac{1}{\sqrt{1-v^{2}/c^{2}}}) \; . \label{82}
\end{equation}

However, by isotropy, the RHS of (\ref{82}) must be independent of $\alpha$. By setting $\alpha= \pi/2$ and subtracting, we have the two
equations,

\begin{equation}
K_{i}(V^{2})-K_{f}(V^{2})=2(u
(\frac{1}{\sqrt{1-V^{2}/c^{2}}\sqrt{1-v^{2}/c^{2}}})-u(\frac{1}{\sqrt{1-v^{2}c^{2}}}))
\; , \label{83}
\end{equation}

and

\begin{equation}
u(\frac{1+vV \cos \alpha/c^{2}}{\sqrt{1-V^{2}/c^{2}}\sqrt{1-v^{2}/c^{2}}})+u(\frac{1-vV \cos
\alpha/c^{2}}{\sqrt{1-V^{2}/c^{2}}\sqrt{1-v^{2}/c^{2}}})\equiv 2u(\frac{1}{\sqrt{1-V^{2}/c^{2}}\sqrt{1-v^{2}/c^{2}}}) \; , \; \textrm{all }
\alpha. \label{84}
\end{equation}

Differentiating (\ref{84}) on $\alpha$, and denoting the arguments of the left hand as $(+)$, $(-)$,

\begin{equation}
u'(+)=u'(-) \; . \label{85}
\end{equation}

Next, since (\ref{84}) is true for any $v$, differentiating on $v$, and substituting (\ref{85}) produces

\begin{equation}
u'(\frac{1+vV \cos \alpha/c^{2}}{\sqrt{1-V^{2}/c^{2}}\sqrt{1-v^{2}/c^{2}}})=u'(\frac{1}{\sqrt{1-V^{2}/c^{2}}\sqrt{1-v^{2}/c^{2}}}) \;
\textrm{all} \; \alpha. \label{86}
\end{equation}

But then $u'$ is independent of its argument over its entire range $\geq1$, hence, $u$ is linear in its argument, and so with the convention
(\ref{79}),

\begin{equation}
k(v^{2})= k_{0}(\frac{1}{\sqrt{1-v^{2}/c^{2}}}-1) \; , \label{87}
\end{equation}

with $k_{0}$ a parameter characterizing the mass of either of the pair. Substituting (\ref{87}) into (\ref{83}), we have

\begin{equation}
K_{i}(V^{2})-K_{f}(V^{2})= \frac{2k_{0}}{\sqrt{1-v^{2}/c^{2}}}(\frac{1}{\sqrt{1-V^{2}/c^{2}}}-1) \; , \label{88}
\end{equation}

so that the K's depend upon $V^{2}$ just as does k. \\

Consider $|v|\ll c$ in (\ref{87}):

$$k(v^{2})= \frac{1}{2}(\frac{k_{0}}{c^{2}})v^{2}+ \ldots\ \; .$$

But, by the experimental \textit{definition} of inertial
mass, the small masses have kinetic energy at low velocities, \\

$$k(v^{2})= \frac{1}{2} mv^{2} \ldots\ \; ,$$
so that

\begin{equation}
k_{0}=mc^{2} \; . \label{89}
\end{equation}

Substituting into (\ref{87}).

\begin{equation}
k(v^{2}) =mc^{2}(\frac{1}{\sqrt{1-v^{2}/c^{2}}}-1) \; , \label{90}
\end{equation}

and $k(v^{2})$ is the correct formula for how kinetic
energy must vary with velocity for $1/c^{2}>0$.  \\

But (\ref{90}) is the formula for the kinetic energy of \textit{any} mass. In particular,

\begin{equation}
K_{i,f}(V^{2})=M_{i,f}c^{2}(\frac{1}{\sqrt{1-V^{2}/c^{2}}}-1) \; . \label{91}
\end{equation}

Comparing (\ref{91}) to (\ref{88}), and using (\ref{89}) and (\ref{90}), we have

\begin{equation}
M_{i} - M_{f} = \frac{2m}{\sqrt{1-v^{2}/c^{2}}}= 2m + \frac{2k(v^{2})}{c^{2}} \; . \label{92}
\end{equation}

(\ref{92}) says that not only did $M_{i}$ weigh more than $M_{f}$ by the emitted mass $2m$, but also weigh more by their
emergent kinetic energy divided by $c^{2}$. \\

Finally, with energy generally conserved, $2k(v^{2})$ is just the energy in the compressed spring while the $m$ pair was still at rest within the
large object. That is,

\begin{equation}
M_{i} - M_{f}= 2m + E_{int}/c^{2} \; . \label{93}
\end{equation}

and the initial object weights more the more the spring is compressed. This is the 1905 result $M=E/c^{2}$, a very different result from
(\ref{90}), or with a different \textit{convention}
$k(0)=mc^{2}$,  \\

\begin{equation}
k(v^{2}) = \frac{mc^{2}}{\sqrt{1-v^{2}/c^{2}}} \; . \label{94}
\end{equation}

(This formula is not $E=mc^{2}$. That formula, (\ref{93}) is about the $m$ of the \textit{numerator} of
(\ref{94}).) \\

So, before and after any rearrangements of particles, \textit{etc.} we know the form of the additive conserved scalar $E$:

\begin{equation}
E= \mathbf\Sigma \mathnormal \frac{mc^{2}}{\sqrt{1-v^{2}/c^{2}}} \; . \label{95}
\end{equation}

Consider (\ref{95}) viewed in world V, using (\ref{75}):

\begin{equation}
E'= \mathbf \Sigma \mathnormal \frac{mc^{2}}{\sqrt{1-v'^{2}/c^{2}}}= \frac{1}{\sqrt{1-V^{2}/c^{2}}}(\mathbf\Sigma \mathnormal
\frac{mc^{2}}{\sqrt{1-v^{2}/c^{2}}}- \textbf{V}\cdot \mathbf\Sigma \mathnormal  \frac{m\textbf{v}}{\sqrt{1-v^{2}/c^{2}}}) \; . \label{96}
\end{equation}

The first quantity within the parenthesis is the conserved $E$. By relativity, $E'$ is also conserved. So by (\ref{96}), we now also know a
conserved rotational vector:

\begin{equation}
\textbf{P} \equiv \mathbf\Sigma \mathnormal \frac{m \textbf{v}}{\sqrt{1-v^{2}/c^{2}}} \; . \label{97}
\end{equation}

Clearly, for small $v$, \textbf{P} is the conserved momentum. It is important to verify that \textbf{P} and $E$ are the correct
\textit{symplectic} momentum and Hamiltonian, after which, if true, we know how to write the kinematics and dynamics for
$1/c^{2}>0$. \\

Consider just one mass, and choose the velocity \textbf{V} in (\ref{96}) to be \textbf{v}, so that $\textbf{v}'=\textbf{0}$:

\begin{equation}
L=\textbf{v} \cdot \textbf{p} - E = -mc^{2}\sqrt{1-v^{2}/c^{2 }} \; ,
 \label{98}
\end{equation}

where if \textbf{p} is the correct momentum for conserved $E$, then
$L$ must be the Lagrangian. \\

But then

\begin{equation}
\textbf{p}=\frac{\partial L}{\partial \textbf{v}}= \frac{m \textbf{v}}{\sqrt{1-v^{2}/c^{2}}} \; , \label{99}
\end{equation}

and so \textbf{P} \textit{is} the mechanical momentum, and the motion of massive particles is determined by the variational principle

\begin{equation}
0= \delta \int L \: dt = - \delta \int mc\sqrt{c^{2}dt- \textbf{dx}^{2}}= - \delta \int mc \: ds \; . \label{100}
\end{equation}

Thus, the package is perfect with $s$ the \textit{invariant} interval of (\ref{74}). From $L$ of (\ref{98}) then follows the Hamiltonian,

\begin{equation}
H_{m} = \sqrt{(c\textbf{p})^{2}+ (mc^{2})^{2}}\; . \label{101}
\end{equation}

Consider now the development of electro\textit{statics} during the $19^{th}$ century. We craft a strong uniform electric field along the
$\hat{\textbf{x}}$-axis, with potential energy, for me at rest with it, for a charge of $+e$,

$$V= -eEx \; .$$

Adding $V$ to $H_{m}$, or alternatively subtracting it from $L$ of (\ref{98}), or alternatively, writing Newton's law for force
$\textrm{\textbf{F}}=eE \hat{\textbf{x}}$ with momentum \textbf{p} of (\ref{99}), we have

\begin{equation}
\frac{d}{dt} \frac{\textrm{\textbf{v}}}{\sqrt{1-v^{2}/c^{2}}}= \frac{e}{m} E \hat{\textbf{x}} \; , \label{102}
\end{equation}

and the ensuing ``hyperbolic" trajectories easily determined, and differing from the parabolic ones of $1/c^{2}=0$. With potential differences of
50kv, electrons move at $\sim c/2$, and the motion measurably different from the parabolas of $1/c^{2}=0$. This was all done in the early 1900's.
Either (\ref{102}) for some $1/c^{2}>0$ would agree with experiment or not. If not, then at last Galileo and his concepts are disproven. Instead,
experiment agrees with (\ref{102}) for some definite value of $1/c^{2}$, incidentally, with $c$ very close to the
measured speed of light.\\

But notice, \textit{none} of the entire package relies on light:
It \textit{is} incidental. \\

\section {A Critique of Light-Based Theories}

Indeed, the special theory of relativity can be developed based on the constancy of the \textit{speed} of light.  Speed, however,
implies the full, independent, rotational invariance of the world of each observer.  That is, this version critically includes the
postulates of isotropy from the beginning, so that all comments about it in this paper need to be understood there.  (It is a quite
technical discussion that one is asking for the constant speed of light, rather than, say, the invariance of a scalar wave equation,
which is a decidedly different request.  To talk of light is to accept Maxwell's equations in empty space.  These tensor equations
must then be \textit{covariant} under allowed transformations, which is the correct discussion, and leads to all, and only those
transformations that simply leave the \textit{speed} invariant. The dimension of space significantly matters in this, and for the
precise form of Maxwell's equations, is true only in three spatial dimensions. This is a very peculiarly detailed technical situation
to regard as the elementary foundational postulates of physical science; these details having emerged only after, from a knowledge of
nothing but light in vacuum, we somehow erected an observable theory of it.) Next this produces not the Lorentz group, but the much
larger conformal group, containing nonlinear transformations, that constitute the full, continuous, invariance group of Maxwell's
equations in empty space, and correspond to a massless theory. Yet, to obtain special relativity, one must excise all but the linear
ones, which then indeed is the Lorentz group, after also deleting the linear dilations. In the deepest sense of meaning, where has
the knowledge been injected that says that this innate invariance is too much? Nonlinears are removed precisely by the postulate of
``homogeneity", requiring the uniform motions of non-light particles, and identical to that part of this paper. (With the now
identified Lie group, the conformal group, by computation one can determine that if any bundle of parallel light rays is to transform
into another bundle of parallel rays, then the nonlinear inversions are forbidden.  This then needs to be an observed property of the
world, where we only know things moving at the speed of light.  It is hard to see how this not just an additional \textit{ad hoc}
technical assumption. Galileo's request for a parallel bundle remaining parallel is his explicit realization that worlds are built of
inertial motions, so that all particles in a parallel bundle of the same speed are at rest to one another. We have technically
determined that the parallel bundle of light can never be construed as a bundle of particles at rest to one another. So, one has no
theoretical nor observational clue as why to impose this. Forgetting all this, one can still say it technically suffices,
to remove the nonlinears, to consider ``homogeneity" restricted to light.)\\

But here the story of light ends.  It is simply impossible to dispatch dilations within a discussion purely based on scale invariant
light.  We now own the full form of \textbf{r}(\textbf{v},\textbf{V}) given by (\ref{71}) and (\ref{72}), which indeed enjoys the
full relativity of (\ref{8}), with a nontrivial \textbf{R} as given by (\ref{69}) with a definite value of $1/c^2$, that of light.
However, rather than (\ref{70}) the right hand sides are free up to an arbitrary multiplicative constant $\lambda$.  Moreover, by
isotropy, we know that $\lambda=\lambda(V^2)$.  Willy-nilly, we have learned much about inertia.  Each system reached by a \textbf{V}
transformation has the property that in it all particles are at rest to one another, and so, the allowed worlds are precisely
Galilean inertial worlds, and the mathematical set of parameters,\textbf{V}, has an obvious physical, experimental meaning.  Indeed,
when I see it at velocity \textbf{V}, it sees me at velocity $-\textbf{V}$, with all timing consistently provided just by light.
Calling the transformation of (\ref{70}) multiplied by the rotational scalar $\lambda(V^2)$, \textbf{T(V)}, we now conclude that its
inverse, back to the space-time of my world is $\textbf{T(-V)}/\lambda^2$.  However, we determine special relativity only when it is
true that $\lambda\equiv 1$. This is precisely what cannot follow just from the world of light, since within its physical laws,
$\lambda$ always cancels identically, whatever its non-0 value is.  Thus, to deduce relativity, we need a postulate to the effect
that $\textbf{T(V)}^{-1}=\textbf{T(-V)}$ for \textit{every} \textbf{V}. But, $|\textbf{V}|<c$, and so any such postulate is now about
inertia.  The requisite postulate has announced itself.  Since \textbf{T}\ depends solely upon \textbf{V}, and we know, via the light
deduction of \textbf{r} that the inverse is to a velocity $\textbf{-V}$, we are requesting that the explicit rule \textbf{T(V)}
applies, unchanged, but for the appropriate value of \textbf{V}, now  $\textbf{-V}$, for the world I see at \textbf{V}.  But with
\textbf{V} arbitrary, this holds for \textit{any} inertial world I see, and so I demand the full principle of relativity for inertia.
However interesting and far the thoughts about just light have gotten us (remarkably far!), the story is only finished with the
additional postulate that all worlds enjoy being equivalent with respect not just to light, but for inertia as well.  (With scale
invariant light, this should have been evident at the outset.) But now we can erase almost all the properties of light from the
deduction, since isotropy, ``homogeneity" and inertial relativity already determines the theory. The only gain we are left with via
this path of deduction is that, in a significantly \textit{ad hoc}, and highly technical manner, we can say that linearity follows
from the weaker assumption of just the parallelism of light rays, rather than for all inertial motions. This, then, is the only
alternative to basing the entire theory on just Galileo, he fully ignorant of almost all properties of light. Epistemologically, the
fragility and brittleness of such a fine detail to have learned from the full theoretical formulation of the theory of light makes
this virtually ludicrous.  Devotees of one spatial dimension should of course know that parallelism of light beams (it's just 1D -
what else?) holds for the general conformal group of transformations, here generated by two arbitrary monotone real functions, so
that here one can only delete nonlinears by recourse to uniform inertial
motions, so that a 1D based story can learn nothing from light.\\

In the end, virtually everything that is required for the non-light theory is required for its version, and almost any iota of
relativity for inertial motion of matter already constitutes enough to have already determined the outcome without light. With just
light, one is left in the curious epistemological position that with time defined just by light, one is barely in the experimental
position to phenomenologically confirm the assumption, yet alone of Maxwell's equations, more specifically.  And even then, with the
Lorentz transformations now imposing properties upon material particles, we are provided with barely any conceptual apparatus about
how to interpret and measure them.  This is the ``alternative" postulate, which has nothing \textit{whatever} to do with isotropy,
but rather the technical issue of what is sufficient to determine the fraudulent hand-waving of ``homogeneity" to determine
linearity.
\\

\section {Historical Perspective And Conclusions}

We have demonstrated that special relativity is fully determined by the development of Galileo's thoughts. Why and how did it take so long for
this to have been realized? We have emphasized that the epistemology of the theory is totally decoupled from any knowledge of the behavior of
light, although \textit{post facto}, it reciprocally can have bearing upon some phenomena
of light. \\

To start, there is some simple history of technologies. Cathode ray tubes with electrons at a significant fraction of $c$ date to the beginning
of the 20th century. (Modern TV tubes are some 50-100kv with electrons at $\sim c/2$. The technology has long required relativistic corrections.)
This technology was developed without electromagnetic fields and the theory of light. The discovery by Michelson of the independence of the speed
of light was preliminary in 1887. Without the unprecedented expense of the experiment, footed by an emerging United States seeking a better
technological and cultural stance, it is unlikely that the failure of Newtonian dynamics would have been first seen in the world of light rather
than in that of cathode rays. It is a close historical accident which came first, although the consistency of electromagnetic theory with its
material sources was driving a light-directed path of inquiry. Had cathode rays determined a violation of Newton,
was someone ready to step in and say, Galileo can be developed also with $1/c^{2} \neq 0$? \\

Why hadn't Galileo determined the full range of systems that embody his thoughts? More seriously, why hadn't the mathematically superior Newton?
Why hadn't the poser of the brachistochrone, Leibniz, nor various of the Bernoulli? More significantly, why hadn't the extraordinarily powerful
Lagrange? And then why not Lorentz? Unbelievably, why not Poincar\'{e}? And why did Einstein need to, and always continued to, base it
inextricably linked to
light? \\

We can make a few guesses. There is a ``villain" in the story, who, of course, is Newton. The reactionary and absolutist treatment in
the ``Principia" (Somewhere, far away, is a place of ``absolute rest", etc.) buried (unquestionably, willfully by the ungenerous
Newton) relativity for almost two centuries. The unprecedented success of astronomy from $1/r^{2}$ enrobed Newton in a posture of
authority that none challenged nor questioned. The first, and exceedingly harsh, challenge came from Mach - but only in 1883. It made
no impact on Lorentz and Poincar\'{e}, who assumed $t'=t$, but that the global, integrable coordinate $t' \neq t$ of the Lorentz
transformations, was merely a ``local", ``internal" time. The observations one made of the world were then locally rearranged by the
matter and fabric of the world to resurrect Newtonian $t'=t$ in an ether at absolute rest. It seems only Einstein came away from Mach
courageous enough to finally and
fully challenge Newton. \\

But then why insist on using light to erect the worlds? Consider that if a part of Newton were to fall, then who could be sure of how
much and what to maintain. Certainly, the neophyte Einstein possessed no such convincing authority. But, the measured word of light
itself did. Perhaps this is why. Who
knows? \\

In conclusion, it is important to know that the foundations of our present kinematics \textit{don't} rely on the properties of light. Should
light, photons, turn out to be composite, they must then acquire some mass. This could be true with $1/c^{2}$ of relativity still being a
limiting constant of Nature, but now without a palpable physical entity directly expressing it. There is no particle that innately is Plank's
constant. It is perhaps worth recalling that the neutrino, crafted as two-component by virtue of its masslessness, seemed necessarily thereafter
to be massless. But, it turns out to have a negligible, but very non-vanishing mass, so that it can no longer be conceived of as moving at $c$.
Surprises are always possible, but they need not overturn, yet, Galileo's brilliant vision. They did so a century
ago for Newton. Such is the pure world of human thought. \\

\section {Comments and Acknowledgements}

This paper was circulated since early September, 2005, and was the subject of seminars I gave in the U.S. and abroad since May, 2005.
I had extensive correspondence with David Mermin for several months, starting in mid-September, 2005, which resulted in some
improvement (or diminution!) of some of my rhetoric, as well as highly useful input that led to my expanding upon several details
that he sensed were too elliptically stated.  I have, as always, deeply enjoyed my discussions with David, and have always garnered
much of assistance.  I here acknowledge his efforts with pleasure.\\

This paper has lain fallow since early 2006, when my interests significantly shifted.  Following some recent new external interest in
the contents of this document, I have returned to it, and added several more, detailed, expansions upon its somewhat elliptical
style.  (This was the traditional style of theoretical physics.  Statements that seem merely intuitive are understood to belie
implicit full rigor.  This seems not to have been understood by some readers, or that I left too much implicit, and so I have fleshed
out the most centrally important such comments.  The reader should be in a position to prove everything that is stated in this paper.
) They are very few in number, and but for a few very brief ones, set off in brackets. One of them is the newly added Section 7, not
set within brackets, as are also the first three paragraphs of Section 2.  Some of them (for example the details of Section 7)
are probably not intelligible to other than cognoscenti.\\

I have decided to let Galileo's Child fly free.  It is my lengthy attestation to the magnificence of Galileo.
\\

\end{document}